\documentclass%
[aps,amsfonts,amssymb,amsmath,superscriptaddress,prc,showpacs,%
letterpaper,twocolumn,nofootinbib]%
{revtex4}
\usepackage[english]{babel}
\usepackage[dvips]{graphicx}

\usepackage{nuclear} % Macros etc. go to nuclear.sty, please

\makeatletter
\def\@dotsep{4.5}
\makeatother

\begin{document}

%%%%%%%%%%%%%%%%%%%%%%%%%%%%%%%%%%%%%%%%%%%%%%%%%%%%%%%%%%%%%%%%%%%%%%%%%%%%%%%

\title{
    Isovector splitting of nucleon effective masses, ab-initio benchmarks and
    extended stability criteria for Skyrme energy functionals
}

%%%%%%%%%%%%%%%%%%%%%%%%%%%%%%%%%%%%%%%%%%%%%%%%%%%%%%%%%%%%%%%%%%%%%%%%%%%%%%%

\author{T. Lesinski}
\email{lesinski@ipnl.in2p3.fr}
\affiliation{
    Institut de Physique Nucl\'eaire de Lyon,
    CNRS-IN2P3/Universit\'e Claude Bernard Lyon 1,\\
    43, bd.~du 11~novembre~1918,
    F-69622 Villeurbanne Cedex, France
}

\author{K. Bennaceur}
\affiliation{
    Institut de Physique Nucl\'eaire de Lyon,
    CNRS-IN2P3/Universit\'e Claude Bernard Lyon 1,\\
    43, bd.~du 11~novembre~1918,
    F-69622 Villeurbanne Cedex, France
}
\affiliation{
    CEA-Saclay DSM/DAPNIA/SPhN, F-91191 Gif-sur-Yvette, France
}

\author{T. Duguet}
\affiliation{
    National Superconducting Cyclotron Laboratory and
    Department of Physics and Astronomy,\\
    Michigan State University, East Lansing, MI 48824, USA
}

\author{J. Meyer}
\affiliation{
    Institut de Physique Nucl\'eaire de Lyon,
    CNRS-IN2P3/Universit\'e Claude Bernard Lyon 1,\\
    43, bd.~du 11~novembre~1918,
    F-69622 Villeurbanne Cedex, France
}

\date{\today}

%%%%%%%%%%%%%%%%%%%%%%%%%%%%%%%%%%%%%%%%%%%%%%%%%%%%%%%%%%%%%%%%%%%%%%%%%%%%%%%

\begin{abstract}
  We study the effect of the splitting of neutron and proton effective masses
  with isospin asymmetry on the properties of the Skyrme energy density
  functional. We discuss the ability of the latter to predict observable of
  infinite matter and finite nuclei, paying particular attention to
  controlling the agreement with ab-initio predictions of the spin-isospin
  content of the nuclear equation of state, as well as diagnosing the onset of
  finite size instabilities, which we find to be of critical importance. We
  show that these various constraints cannot be simultaneously fulfilled by
  the standard Skyrme force, calling at least for an extension of its
  P-wave part.
\end{abstract}

\pacs{
    21.30.Fe, % Forces in hadronic systems and effective interactions
    21.60.Jz  % Hartree-Fock and random-phase approximations
}

%%%%%%%%%%%%%%%%%%%%%%%%%%%%%%%%%%%%%%%%%%%%%%%%%%%%%%%%%%%%%%%%%%%%%%%%%%%%%%%
%%%%%%%%%%%%%%%%%%%%%%%%%%%%%%%%%%%%%%%%%%%%%%%%%%%%%%%%%%%%%%%%%%%%%%%%%%%%%%%

\maketitle

%%%%%%%%%%%%%%%%%%%%%%%%%%%%%%%%%%%%%%%%%%%%%%%%%%%%%%%%%%%%%%%%%%%%%%%%%%%%%%%

\section{Introduction}

%%%%%%%%%%%%%%%%%%%%%%%%%%%%%%%%%%%%%%%%%%%%%%%%%%%%%%%%%%%%%%%%%%%%%%%%%%%%%%%

In the study of medium to heavy mass nuclei, nuclear Energy Density Functional
(EDF) approaches, based on self-consistent Hartree-Fock (HF)
methods and their extensions, constitute the theoretical tool of
choice~\cite{bender03b}. Thanks to the
development of better energy functionals and to the increase of computer
resources, nuclear EDF is on the edge of becoming a predictive theory for all
nuclei but the lightest. This is not only true for ground state properties,
such as binding energies, radii or multipoles of the density, but also for
low-energy spectroscopy and decay probabilities~\cite{bender03b}.

However, the accuracy and predictive power needed for unknown regions of the
nuclear chart still leave a lot of room for improvement. The phenomenological
nature of Skyrme functionals makes their ability to faithfully predict
observable or phenomena not linked with those used for their construction quite
weak. Indeed, the limited number of adjustable parameters (compared to the
wealth of nuclear observable to be matched) turns fitting a Skyrme functional
into an overconstrained problem (which, of course, does not prevent some parts
of it from being underconstrained).

As a direct consequence, many properties of existing parameterizations
are biased to the fitting procedure and the limited analytical form of
the Skyrme force, rather than to physical reasoning. A well-known example
is the equation of state (EOS) of Pure Neutron Matter (PNM), which is sometimes
subject to a pathological collapse at high density when not explicitly
constrained. This is problematic insofar as one of the major challenges of
contemporary nuclear theory is to predict properties of very isospin asymmetric
nuclear systems, \emph{i.e.~}neutron rich nuclei and matter in neutron stars.
Experimental data being unavailable in this domain of isospin, one has started
relying on ab-initio theoretical results to constrain isovector properties of
the functional. It has led to the construction of the ``Saclay-Lyon'' SLy
series of parameterizations \cite{chabanat97,chabanat98,chabanat98b} by
fitting (among other quantities) a theoretical equation of state of neutron
matter.

Isovector features of the nuclear EOS are crucial for a good
understanding of neutron stars, exotic nuclear collisions produced at
radioactive beam facilities and to describe the structure of exotic nuclei. For
instance, the density dependence of the volume symmetry energy determines the
proton fraction in $\beta$ equilibrium in neutron stars, which ultimately
drives the cooling rate and neutrino emission~\cite{lattimer04}. The
high-density part of the symmetry energy, which happens to be strongly model
dependent, also influences significantly the isospin diffusion in heavy-ion
collisions~\cite{chen04a}. Finally, the low-density part of the symmetry energy
is correlated with the size of neutron skins in finite nuclei~\cite{typel01}.

Beyond global isospin-dependent properties of the EOS, the isovector part of
nucleon-dependent quantities may influence the behavior of the above mentioned
systems. Thus, collision observable depend on the momentum dependence of the
mean-field, in particular on its isovector component~\cite{li04,li04b}. Also,
some properties of neutron stars require a precise knowledge of isoscalar and
isovector nucleon effective~masses~\cite{bethe90a,farine01a}. The latter, which
drives the \emph{splitting of neutron and proton effective masses} with
neutron/proton asymmetry, will serve as a starting point for our study. Indeed,
a lot of efforts has recently been devoted to the microscopic characterization
of neutron and proton effective masses in infinite Asymmetric Nuclear Matter
(ANM)~\cite{bombaci91,kubis97,lombardo,greco01,hofmann01,liu02,rizzo04,ma04a,
dalen05,satula06}.
Either in ANM or in nuclei, the two species acquire different effective masses.
This property is quantified by the difference $\dmi = \emn(I) - \emp(I)$, where
$I=(\rhon-\rhop) / (\rhon+\rhop)$ is the isospin asymmetry while $\rhon$ and
$\rhop$ denote neutron and proton densities, respectively. Note that the
different effective masses $m^{\ast}$ discussed in the text always refer in
fact to the ratio $m^{\ast}/m$, where $m$ is the bare nucleon mass. The latter
is taken to be the same for neutrons and protons.

This effective-mass splitting, though, is only one of a wealth of observable
which can be subject to comparison between ab-initio predictions and EDF
models. In this work we present results of a classical yet long unused test:
the separation of infinite Symmetric Nuclear Matter (SNM) potential energy per
particle into spin-isospin channels.

We shall also pay particular attention to controlling instabilities
(\emph{i.e.}~non-physical spontaneous breaking of spin, isospin and/or spatial
symmetries), and correlate $\dmi$ with \emph{vector} properties of the
functional. We thus investigate the behavior of the latter with respect to the
breaking of time-reversal invariance and the onset of spin polarization,
looking for an overall consistency check of its spin-isospin content. Indeed,
such properties will become more and more important as one attempts to use
full-fledged Skyrme functionals to study odd-mass nuclei, calculate
rotational properties through self-consistent cranking calculations, or use
more general dynamical methods~\cite{bender02}.

This paper is organized as follows: in \citesection{effectivemass} we present
the set of Skyrme parameterizations used in this study and examine basic
properties of nuclear matter and finite nuclei. From then,
in~\citesection{moreinm} we perform a more detailed study of the spin-isospin
content of the functionals and of their stability against finite-size spin
and isospin perturbations using response functions in the Random Phase
Approximation (RPA).

%%%%%%%%%%%%%%%%%%%%%%%%%%%%%%%%%%%%%%%%%%%%%%%%%%%%%%%%%%%%%%%%%%%%%%%%%%%%%%%

\section{Constraining the isovector effective mass}
\label{sec:effectivemass}

%%%%%%%%%%%%%%%%%%%%%%%%%%%%%%%%%%%%%%%%%%%%%%%%%%%%%%%%%%%%%%%%%%%%%%%%%%%%%%%

The nucleon effective mass $m^{\ast}$ is a key property
characterizing the propagation of (quasi)nucleons through the
nuclear medium~\cite{jeukenne76a}. It is a reminder of the
non-locality and energy dependence of the nucleon self-energy
$\Sigma (k,\epsilon)$, themselves originating from the finite
range and non-locality in time and space of the nucleon-nucleon
interaction. Mean-field-like theories of finite nuclei or infinite
matter assume an on-shell propagation of the nucleons. The
effective mass associated with such an on-shell propagation does
not take into account the fragmentation of the single-particle
strength and thus, a limited part of the effects associated with
the energy dependence of $\Sigma (k,\epsilon)$. Finally, the total
energy is calculated by considering the quasi-holes (particles) to
have spectroscopic factors of 1. In this context, either
microscopic~\cite{baldo99} or making use of phenomenological
interactions or functionals~\cite{bender03b}, mean-field methods
do not correspond to a naive Hartree-Fock theory and always amount
to renormalizing a certain class of correlations into the
effective vertex. However, the energy dependence of the
self-energy arising from the correlations only influences the
position of the quasi-particle peak energy. The standard nuclear
EDF methods differ in several respects from Kohn-Sham Density
Functional Theory~\cite{kohn64a}. In particular, the strategy
usually followed in the former is to leave room for improvement
through further inclusions of correlations. This can be done by
performing (Quasiparticle)-Random-Phase-Approximation (QRPA)
calculations~\cite{blaizot77,sev02} or by employing the Generator
Coordinate Method (GCM) on top of symmetry-restored mean-field
states~\cite{meyer95}.

Thus, the effective mass adjusted at the pure mean-field level is not
expected
to generate single-particle spectra matching exactly experimental
data extracted from neighboring odd-mass nuclei. In particular,
the coupling of single-particle motions to surface vibrations in
closed-shell nuclei is known to increase the density of states at
the Fermi surface and thus the effective
mass~\cite{bernard80a,goriely03}. A mean-field isoscalar effective
mass $\ems$ lying in the interval $0.7/0.8$ in SNM, is able to
account for a good reproduction of both isoscalar quadrupole giant
resonances data in doubly closed-shell nuclei~\cite{liu76a} and of
single-particle spectra in neighboring ones provided
particle-vibration coupling has been properly included. When the
latter coupling is taken into account, the effective mass becomes
greater than one for states near the Fermi surface. Certainly, a
lot remains to be done to understand these features
microscopically in more involved cases~\cite{charity06}. This is
not only true for mid-shell nuclei where the coupling to both
rotational and vibrational states can be important, but also for
exotic nuclei where the coupling to the continuum becomes crucial
and where shape coexistence and/or large amplitude motion appear
more systematically.

In very exotic systems, the isovector behavior of $\emp$ and $\emn$ should
play
an important role. However, so far, no experimental data from finite nuclei has
allowed a determination of the effective mass splitting as a function of
neutron richness. In this context, ab-initio calculations of ANM are
of great help. Non-relativistic Brueckner-Hartree-Fock (BHF) calculations, with
or without three-body force, and, with or without rearrangement terms in the
self-energy, predicted $\Delta m^\ast(I)$ to be such that $\emn \geq \emp$ in
neutron-rich matter, that is, for $I \geq 0$. Such a conclusion was also
reached
by calculating the energy dependence of the symmetry potential (the Lane
potential~\cite{lane62}) within a phenomenological formalism~\cite{li04}.
The latter result was confirmed by microscopic Dirac-Brueckner-Hartree-Fock
(DBHF) calculations~\cite{sammarruca04a}. The situation regarding the
prediction of the
effective mass splitting was complexified due to an apparent contradiction
between results obtained from BHF~\cite{bombaci91,lombardo} and DBHF
calculations~\cite{hofmann01}. However, the situation was finally clarified in
Ref.~\cite{ma04a,dalen05} where the importance of the energy dependence of the
self-energy and the need to compare the non-relativistic effective mass with
the vector effective mass in the relativistic framework~\cite{jaminon89} were
pointed out.

Thus, the sign of the splitting is rather solidly predicted. However, its
amplitude is subject to a much greater uncertainty. Starting from that
observation, the goal of the present section is to study the impact of
the effective-mass splitting on properties of exotic nuclei predicted by
Skyrme-Hartree-Fock calculations. As far as the effective-mass splitting is
concerned, one expects consequences onto structure properties of neutron-rich
nuclei. As a relatively large asymmetry may be necessary to reveal the
influence
of the splitting, data from nuclei not yet studied experimentally should
provide crucial information in that respect. As the effective mass governs
the density of states at the Fermi surface (together with the spin-orbit and
the tensor forces), the amplitude of the splitting may influence properties
such as masses and single particle properties of exotic nuclei, the evolution
of
isotopic shifts across neutron rich closed-shell nuclei or shell corrections in
superheavy nuclei around the $(N=184,Z=120)$ island of
stability~\cite{bender_she99,kruppa,naza_she02,bitaud}. Also, neutron and
proton
correlations beyond the mean-field should develop rather differently depending
on the direction and amplitude of the effective-mass splitting. This could be
true for static and dynamical pairing correlations as well as for the coupling
to vibrational and rotational states. Finally, the effective mass splitting
should leave its fingerprint onto the characteristics of isovector vibrational
states of different sorts in neutron rich
nuclei~\cite{paar05}.

%=============================================================================%

\subsection{Fitting protocol}
\label{sec:fit}

%=============================================================================%

Trying to keep a coherence in the way we construct Skyrme functionals, we take
the fitting protocol used to define the SLy
forces~\cite{chabanat97,chabanat98,chabanat98b}
as a basis for the present work. Also, we pay attention to the fact that any
improved or complexified functional includes all features validated by the SLy
ones.

The SLy functionals were derived from an effective interaction, that is,
time-odd components of the functional are eventually linked to time-even ones.
However, some terms in the functional given by
Eqs.~\ref{eq:skfunc1}-\ref{eq:skfunc4} were dropped for some of the
parameterizations. For instance, time-odd terms of the form $\svec_{q} \cdot
\triangle \svec_{q'}$ have not been considered when calculating odd
nuclei~\cite{duguet02a} or rotational states~\cite{rigol}. Also, $\Jtens^{2}$
contributions to the energy functional associated with momentum-dependent terms
in the central Skyrme force have been omitted for some of the SLy
parameterizations, as it has been the case for most of the Skyrme
functionals so far~\cite{bender03b}. Rigorously, omitting terms is inconsistent
with the idea of deriving a functional from an interaction. In any case,
the latter approach is only used for simplicity until proper adjustment or
derivation
of time-odd terms is feasible. We refer the reader to
Refs.~\cite{chabanat97,chabanat98,chabanat98b,bender03b} for a more extensive
discussion on the subject.
We presently take the SLy5 parameterization as a starting point. Thus, the
two-body part of the center of mass correction is omitted whereas the
$\Jtens^{2}$ terms are fully kept. The spin-orbit term is the standard one,
with a single parameter adjusted on the splitting of the $3p$ neutron
level in \nuc{208}{Pb}.

Within this general scheme, we have built a series of three new Skyrme forces,
denoted hereafter $f_-, f_0$ and $f_+$. The departures from the SLy protocol
considered presently are (i)~a better control of spin-isospin instabilities via
Landau parameters (ii)~the use of two density-dependent zero-range
terms~\cite{cochet04a} (iii)~a constraint on the isovector effective mass, such
that, in neutron-rich systems, $\emn<\emp$ for $f_-$, $\emn=\emp$ for $f_0$ and
$\emn>\emp$ for $f_+$.

With two density dependent terms, the compressibility and the isoscalar
effective mass are no longer bound together and can be chosen independently.
However, this is not directly used here and an isoscalar effective mass of
$\ems = 0.7$, close to the SLy5 value, is chosen for the three
parameterizations $f_-,
f_0, f_+$. The additional freedom brought about by the second density-dependent
term is only used to adjust more easily the high-density part of the PNM EOS
(see below). In the end, the only parameter subject to variation between $f_-$,
$f_0$ and $f_+$ is the isovector effective mass $\emv$ which, $\ems$ being
constant, drives the splitting $\Delta m^\ast(I)$.

In the present work, we use the SLy5 force as a reference, and include a
comparison with the recently proposed LNS parameterization~\cite{cao06a}
which was also built to match the splitting of effective masses and the neutron
matter EOS predicted by BHF calculations. The SkP
force~\cite{doba84a}, initially built for the study of pairing
effects, will be used for a special purpose in the discussion about
instabilities.

%=============================================================================%

\subsection{Elementary properties of studied forces}

%=============================================================================%

As we focus on the behavior of effective masses $\emq$ with isospin
asymmetry, we recall that these quantities are related to the dependence of the
energy density functional, Eqs.~\ref{eq:skfunc1}--\ref{eq:skfunc4}, on
kinetic densities $\tau_q$, as
\begin{subeqnarray}
\frac{\hbar^2}{2\emq(I)}
    &=& \frac{\partial \CH}{\partial \tau_q}
    ~=~ \frac{\hbar^2}{2m} + \ctau_0~ \rhoisos + qI~\ctau_1~\rhoisos\\
\frac{m}{\emq(I)} &\equiv& \frac{m}{\ems}
		+ qI~\left(\frac{m}{\ems} - \frac{m}{\emv} \right)
    \label{eq:xmstarq}
\end{subeqnarray}
where $\rhoisos$ is the scalar-isoscalar density
and $q=+1,-1$ respectively for neutrons and protons (for the definition of
$\ctau_t$ coefficients, see \citeappendix{dft}). The splitting of effective
masses, quantified by
\begin{eqnarray}
    \frac{\dmi}{m} &=& \frac{\emn(I)}{m} - \frac{\emp(I)}{m},
\end{eqnarray}
is governed by the isoscalar and isovector effective masses
\begin{subeqnarray}
    \frac{m}{\ems} ~=& 1 + \frac{2m}{\hbar^2}~\ctau_0~\rhoisos
        &\equiv~ 1 + \kaps, \\
    \frac{m}{\emv} ~=& 1 + \frac{2m}{\hbar^2}~\left(\ctau_0-\ctau_1\right)
         ~\rhoisos
        &\equiv~ 1 + \kapv.
\end{subeqnarray}
We use the usual convention for the isovector effective mass, which stems from
its definition through the enhancement factor $\kapv$ of the Thomas-Reiche-Kuhn
sum rule~\cite{trk}. However, $\emv$ and $\kapv$ are \emph{not} isovector
quantities in the sense of isovector couplings of the functional.

In the following, we shall discuss the value of $\dmi$ at $I=1$, which we note
$\dmiun$ in the following, for the sake of brevity. We have
\begin{eqnarray}
    \frac{\dmiun}{m}
        &=& \frac{2(\kapv - \kaps)}{(1 + \kaps)^2 - (\kapv - \kaps)^2},
\end{eqnarray}
such that $\dmiun > 0$ for $\kapv > \kaps$, or equivalently $\emv < \ems$, or
$\ctau_1 < 0$.

%.............................................................................%

\newcommand{\tc}[1]{\multicolumn{2}{c}{#1}}
\newcommand{\hsa}{\hspace{0.6em}}
\newcommand{\hsb}{\hspace{1.0em}}
\newcommand{\hsc}{\hspace{1.6em}}
\begin{table}[htbp]
    \caption{Infinite nuclear matter properties of the Skyrme forces quoted
        in the text. The quantities $\rhosat$ and $E/A$ denote the density and
        energy per particle at saturation in SNM. The symmetry energy and
        the compressibility (for symmetric matter) are respectively
        32~MeV and 230~MeV for SLy5 and all $f_x$ forces. In the case
        where $\ems \sim 0.7$, $\kaps\sim 0.43$, so we have $\dmiun>0$
        if $\kapv \gtrsim 0.43$.}
    \smallskip
    \label{tab:propf}
    \begin{tabular*}{\columnwidth}{l*{6}{@{\hfil\hspace{0.82em}}r@{}l}}
        \hline\hline
        Force \hfil&
        \tc{$\rhosat$} & \tc{$E/A$} & \tc{$\ems$} & \tc{$\kapv$} &
        \tc{$\emv$} & \tc{$\dmiun$} \\
        \hline
        SLy5   &\hsa 0&.161 & -15&.987 & 0&.697 & 0&.25 & 0&.800 & -0&.182 \\
        \hline
        $f_-$  & 0&.162 & -16&.029 & 0&.700 &\hsa 0&.15 & 0&.870 & -0&.284 \\
        $f_0$  & 0&.162 & -16&.035 & 0&.700 & 0&.43 &\hsa 0&.700 &  0&.001 \\
        $f_+$  & 0&.162 & -16&.036 & 0&.700 & 0&.60 & 0&.625 &\hsc  0&.170 \\
        \hline
        LNS    & 0&.175 &\hsa -15&.320 & 0&.825 & 0&.38 & 0&.727 &  0&.227 \\
        SkP    & 0&.170 & -16&.590 &\hsa 1&.030 & 0&.32 & 0&.760 &  0&.418 \\
        \hline\hline
    \end{tabular*}
\end{table}

%.............................................................................%

Bulk properties of $f_x$ parameterizations are displayed in \citetable{propf}.
We
note that, while the position of the saturation point varies little between our
forces (SLy5 and $f_x$), this consistency is lost in the case of LNS and SkP.
These properties depend on the observable used in the fitting procedure. In the
case of LNS, the saturation point relates to an Extended
Brueckner-Hartree-Fock (EBHF) calculation~\cite{lombardo}, predicting values of
$(E/A)_\mathrm{sat}$ and $\rhosat$ which are larger than empirical ones. A
similar but lesser trend is observed for SkP. In this case it seems to be
correlated with the choice of effective masses and their interplay with other
parameters of the force. Indeed, binding energies computed with SkP compare
satisfactorily with experimental ones, while LNS suffers in this respect from
the
lack of readjustment of the saturation point on nuclear data. As it has been
shown in Ref.~\cite{bertsch05a}, nuclear binding energies are highly sensitive
to the choice of the energy at saturation, which is therefore constrained to a
very tight interval if one wants to reproduce such quantities. This constraint
is especially tight compared to the uncertainty of ab-initio predictions.
Despite the fit of surface properties ($\cdrho_0$ parameter) on a set of
nuclear
data, the accuracy of binding energies predicted by LNS is of the order of 5\%,
to be compared with less than 1\% for SLy5.

%=============================================================================%

\subsection{Properties of the nuclear matter EOS}

%=============================================================================%

It is interesting to note that SLy parameterizations were fitted to PNM EOS
with the idea of improving isospin properties of the functionals. One
consequence was to generate functionals with $\dmiun<0$, in opposition to
ab-initio predictions. On the other hand, older functionals such as
SIII~\cite{beiner75a} and SkM$^\ast$~\cite{bartel82a}, which were not fitted
to PNM, had $\dmiun>0$. The same exact situation happens for the Gogny
force~\cite{chappert06a}. Thus, improving global isovector properties (EOS)
seems to deteriorate those related to single-particle states ($\emv$) with
currently used functionals. This can be better understood by examining the
expressions for SNM and PNM EOS:
\begin{subeqnarray}
    \frac{E}{A}(\rhoisos, I=0)
        &=& \frac{3}{5} \frac{\hbar^2}{2m}
             \left(\frac{3\pi^2}{2}\right)^{2/3} \rhoisos^{2/3}
                                          + \crho_0(\rhoisos)~\rhoisos
            \nonumber \\
        && +~\ctau_0 \frac{3}{5} \left(\frac{3\pi^2}{2}\right)^{2/3}
                                                       \rhoisos^{5/3}, \\
    \frac{E}{A}(\rhoisos, I=1)
        &=& \frac{3}{5} \frac{\hbar^2}{2m} \left(3\pi^2\right)^{2/3}
                                               \rhoisos^{2/3} \nonumber\\
        &&    + ~[\crho_0(\rhoisos) + \crho_1(\rhoisos)]\rhoisos
            \nonumber \\
        && +~[\ctau_0 + \ctau_1]~\frac{3}{5} \left(3\pi^2\right)^{2/3}
                                               \rhoisos^{5/3}\,.
    \label{eq:pnmeos}
\end{subeqnarray}

If $\crho_t(\rho_0)$ coefficients only contain one low power of the density
($\propto \rhoisos^{1/6}$), the latter influences low-density parts of the EOS
more than high-density ones. The effective mass term then determines the
high-density part of the EOS. In SNM, this translates into the well-known
relation between $\ems$ and the incompressibility
$K_\infty$~\cite{chabanat97,chabanat98,chabanat98b}. In
the case of PNM, the EOS above $\rhosat$ is then mostly fixed by the term
proportional to $\ctau_0 + \ctau_1$ in \citeeqdot{pnmeos}, and any attempt to
use the density dependence to counteract its effects, results in a very strong
constraint on the latter. This in turn degrades the behavior of the functional
at and below saturation density and the fit to properties of finite nuclei. We
recall at this point that the condition $\dmiun > 0$ corresponds to $\ctau_1 <
0$, which drives the high-density PNM EOS down and explains why usual Skyrme
functionals predict either a collapse of the PNM EOS if $\dmiun>0$, or, like
the SLy functionals fitted to PNM EOS, the wrong sign of the effective mass
splitting in neutron rich matter.

If $\crho_t(\rho_0)$ coefficients contain an additional density dependence
with a higher
power, the previous discussion does not apply: using two density-dependent
terms in the functional ($\propto \rhoisos^{1/3}; \rhoisos^{2/3}$)~
\cite{cochet04a} allowed us to construct ($f_{-}$, $f_{0}$, $f_{+}$) with a
good fit to PNM EOS, a free choice of effective masses and satisfactory nuclear
properties.

\begin{figure}[htbp]
    \centering
    \includegraphics[width=0.8\columnwidth]{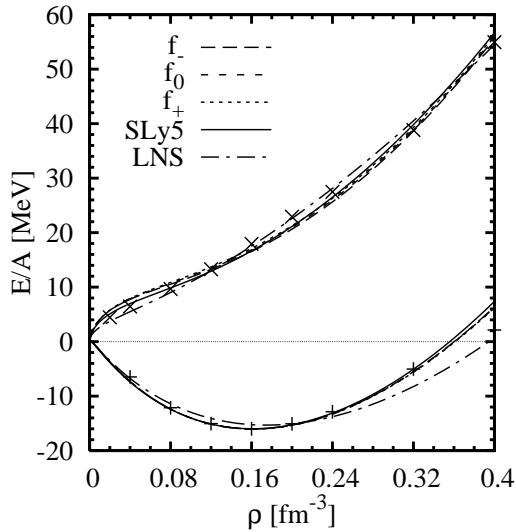}
    \caption{SNM and PNM EOS as given by Skyrme functionals
    presently discussed (see text), compared with VCS results by
    Akmal~\emph{et~al.}~\cite{akmal98a} (crosses: PNM, plusses: SNM).}
    \label{fig:eos}
\end{figure}

The previous discussion already shows the type of problems and information
arising from our attempt to improve on the fitting protocol of SLy functionals
by using more inputs from ab-initio calculations.
Now, Fig.~\ref{fig:eos} shows SNM and PNM EOS as obtained from ($f_{-}$,
$f_{0}$, $f_{+}$, SLy5) and as predicted by Variational Chain Summation (VCS)
methods~\cite{akmal98a}. At this point, one can see that the four forces
($f_{-}$, $f_{0}$, $f_{+}$, SLy5) reproduce both microscopic EOS with the same
accuracy. However, it remains to be seen whether or not this translates into
identical global spin-isospin properties and into similar nuclear structure
properties.

%.............................................................................%

%.............................................................................%

%=============================================================================%

\subsection{Effects on properties of nuclei}

%=============================================================================%

We now study the effects of the variation of the isovector effective mass on
selected properties of spherical nuclei. We start with HF single-particle
energies, then binding energies, ending with a short sum-rule based analysis
of isovector giant resonances.

For computations of open-shell nuclei, we use, in the particle-particle
channel, a zero-range interaction with a density dependent (mixed
surface and volume) form~factor defined~as:
\begin{equation}
	V_\mathrm{pair}(\mathbf R,\mathbf r) = V_0~\delta(\mathbf{r})
		\left[ 1 - \frac{\rhoisos(\mathbf R)}{2\rhosat} \right]\,,
\end{equation}
where $\mathbf R=(\mathbf r_1+\mathbf r_2)/2$ and
$\mathbf r=\mathbf r_1-\mathbf r_2$.
The local HFB equations are renormalized following the procedure developed in
Ref.~\cite{bulgac}.

The strength $V_0$ is adjusted to the mean pairing gaps of six semi-magic
nuclei
(neutron gaps in \nuc{120}{Sn}, \nuc{198}{Pb}, \nuc{212}{Pb} and proton gaps in
\nuc{92}{Mo}, \nuc{144}{Sm} and \nuc{212}{Rn}). In this procedure we compute
theoretical spectral gaps defined as
$\Delta_{th} = \mathrm{Tr}(\tilde{h}\tilde{\rho})/\mathrm{Tr}(\tilde{\rho})$,
where $\tilde h$ is the pairing field and $\tilde\rho$ the pairing
density~\cite{doba84a},
and adjust
each of them upon an experimental gap extracted through a five point difference
formula from masses of neighboring nuclei, as suggested in
Ref.~\cite{duguet02b}.

%---------------------------------------%

\subsubsection{Single-particle energies}

%---------------------------------------%

Effective masses are known to control the average density of single-particle
states. It is thus interesting to check to what extent such statement applies
to neutron-rich nuclei when varying $\emv$.
In this part of the study, we are mainly interested in evaluating
the change in the single-particle energies generated by the functional
for different splittings and not directly by a comparison with
experimental results.

Single-particle energies in \nuc{132}{Sn} and \nuc{208}{Pb} are plotted on
\citefigdot{spe1}. The general trend followed by neutron states with increasing
$\dmiun$ (from force $f_-$ to $f_+$) corresponds to an increase of the
density of neutron states: they tend to come closer to the Fermi energy
$\varepsilon_\mathrm{F}$;
notable exceptions being both
neutron $1i$ levels in \nuc{208}{Pb}. The opposite behavior is observed in
proton levels, which spread away from $\varepsilon_\mathrm{F}$ with increasing
$\dmiun$ (except for
the proton $1h_{11/2}$ level). However, these trends are rather marginal, which
can be linked with the moderate bulk asymmetry of these nuclei ($I = (N-Z)/A =
0.24$ for \nuc{132}{Sn} and $0.21$ for \nuc{208}{Pb}). This moderate asymmetry
means that the isovector term in the definition of the effective mass
(\citeeqdot{xmstarq}) is weakly probed.

\begin{figure}[htbp]
  \centering
  \includegraphics[width=.49\columnwidth]{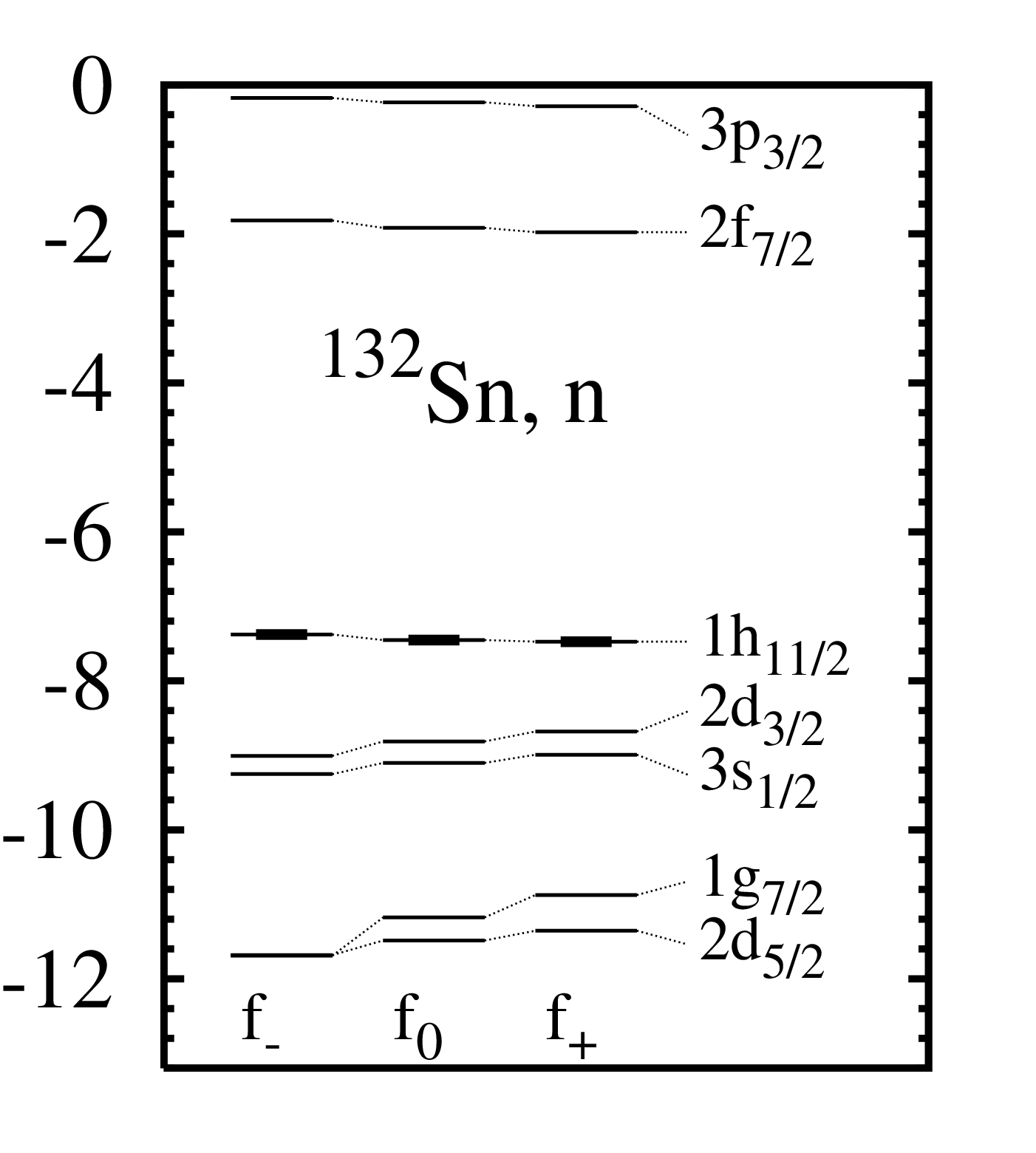}
  \includegraphics[width=.49\columnwidth]{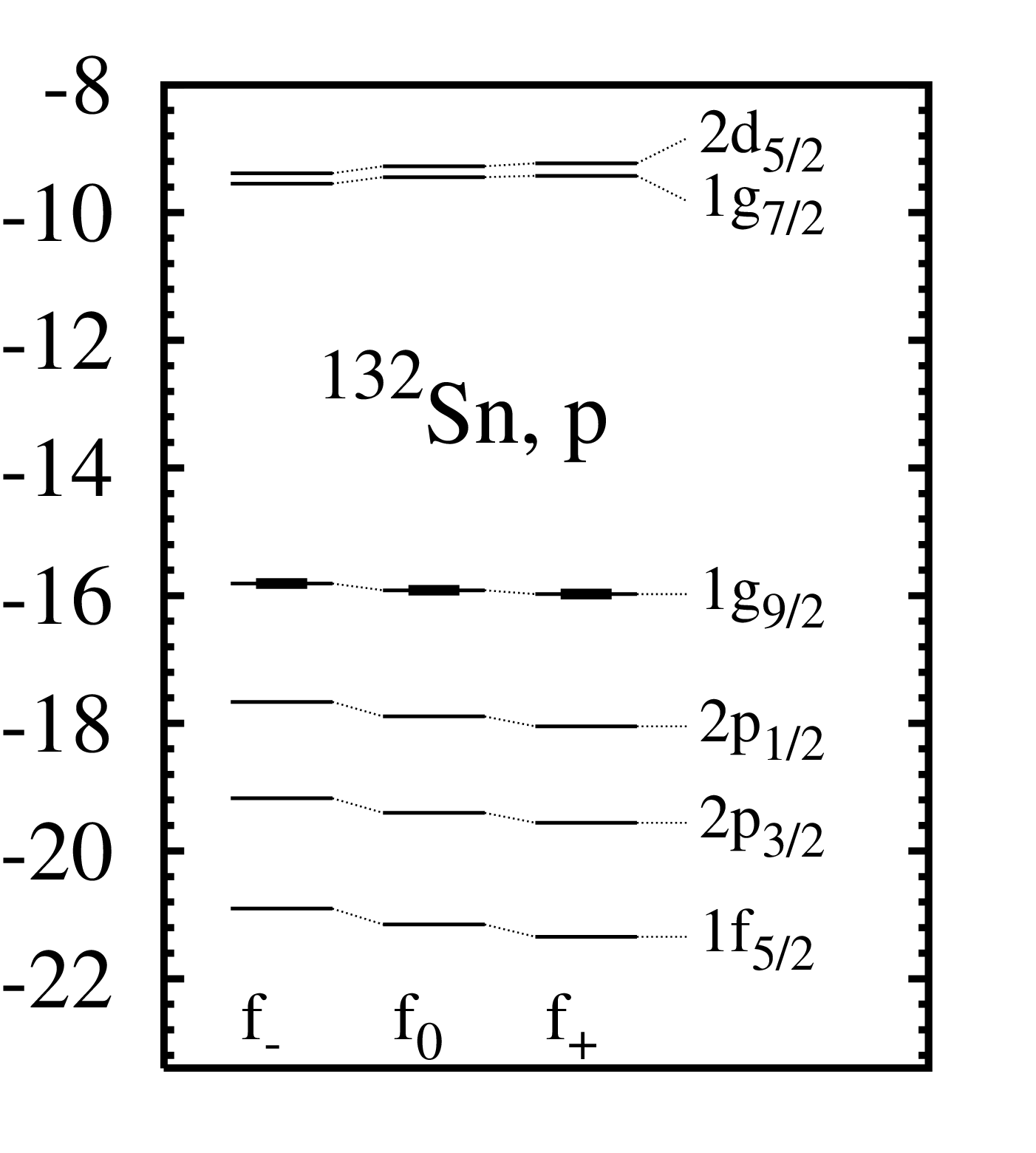} \\
  \includegraphics[width=.49\columnwidth]{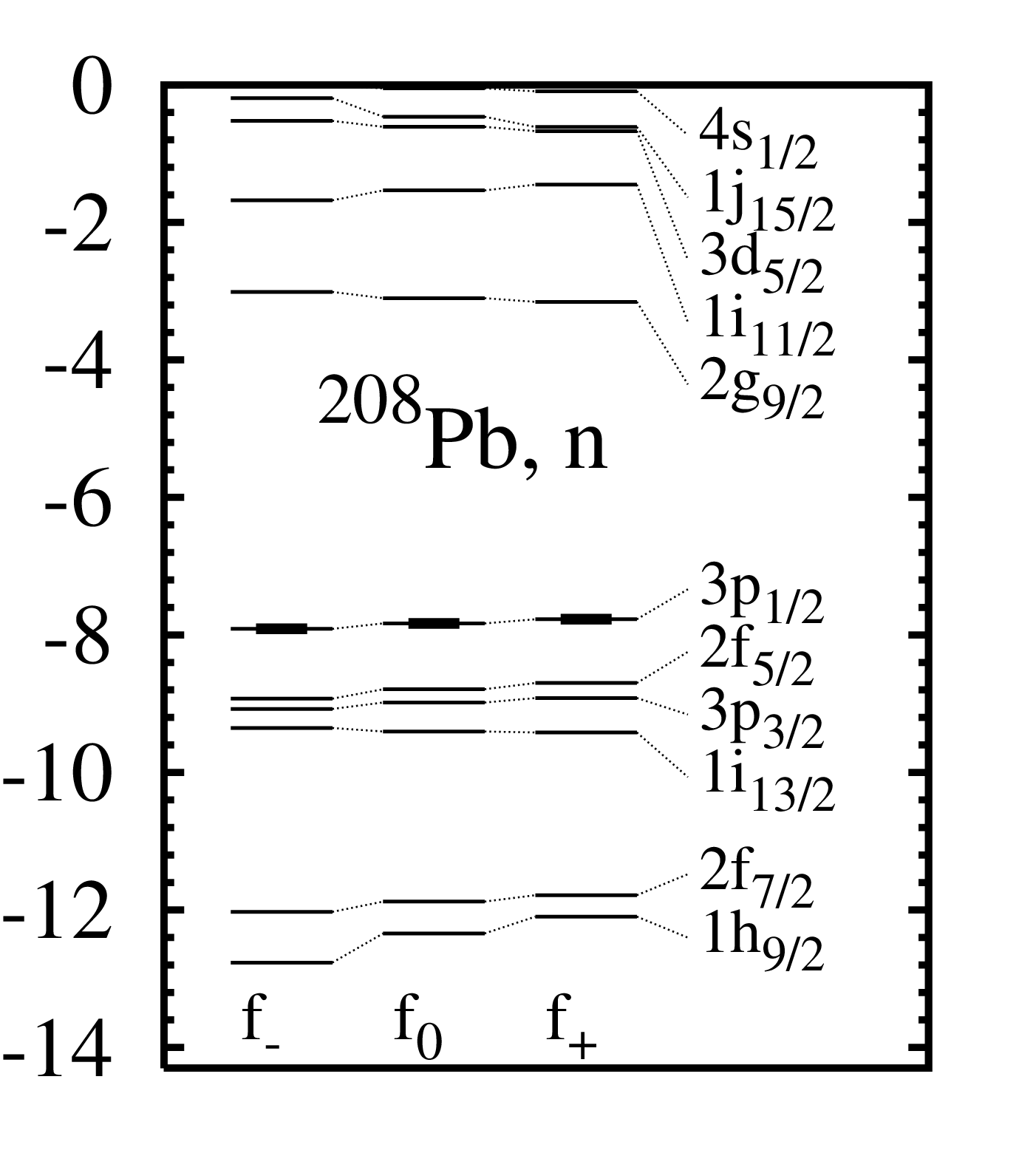}
  \includegraphics[width=.49\columnwidth]{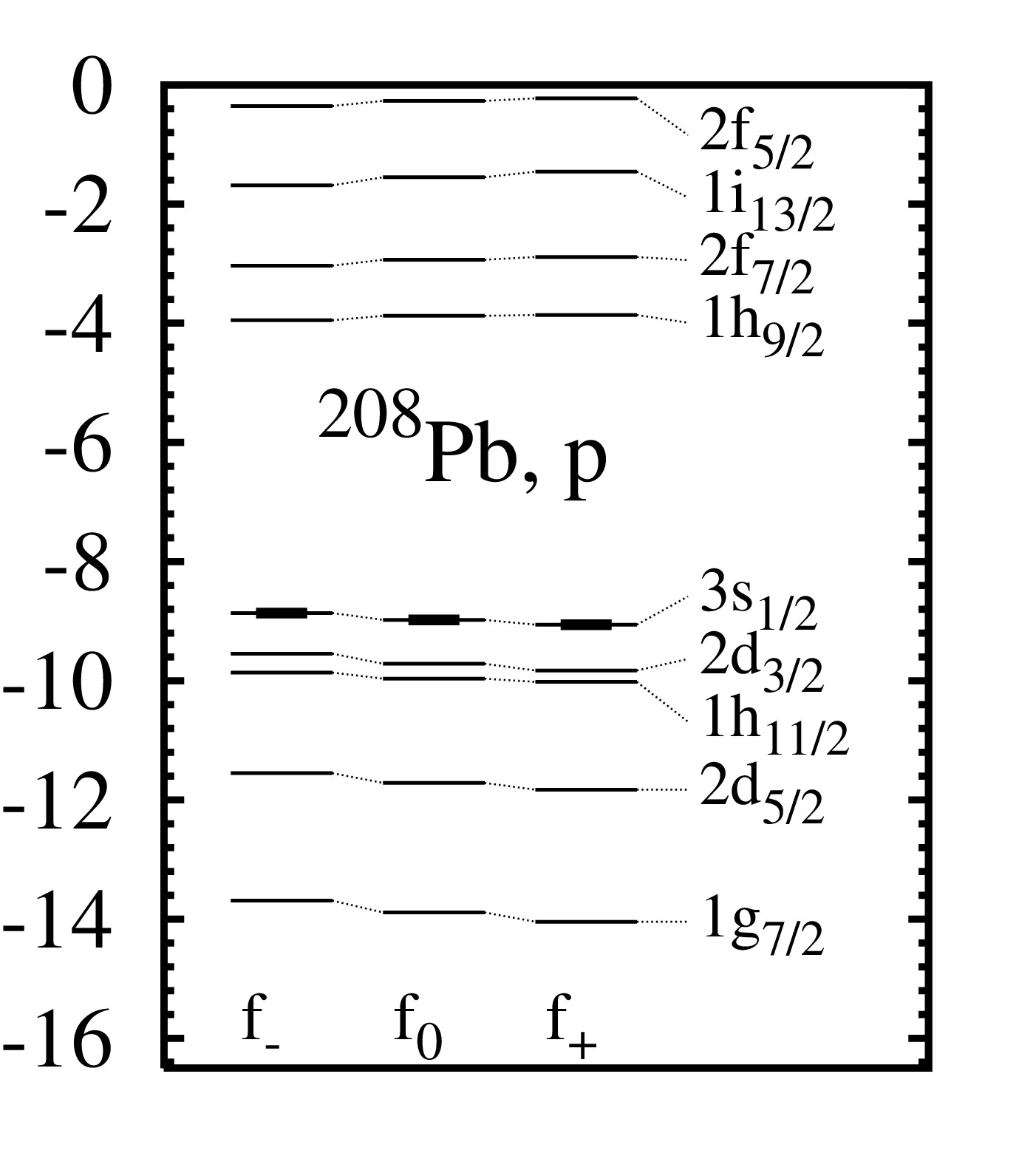} \\
  \caption{Single-particle energies [MeV] in \nuc{132}{Sn} and \nuc{208}{Pb}
           computed with indicated forces. Thick lines indicate the Fermi level
           $\varepsilon_F$.}
           \label{fig:spe1}
\end{figure}

Let us therefore examine similar spectra for more neutron-rich nuclei,
\emph{i.e.~}
\nuc{78}{Ni} ($I=0.28$, experimentally observed~\cite{hosmer}) and
\nuc{156}{Sn}
($I=0.36$). The \nuc{156}{Sn} nucleus is used as an example of an extremely
asymmetric system, even beyond the reach of planned radioactive beam
facilities~\cite{sp2}. We observe on the lower right panel of \citefigdot{spe2}
that the effect of $\Delta m^\ast$ on proton single-particle energies at $Z=50$
is more pronounced
in \nuc{156}{Sn} than it was in \nuc{132}{Sn}. The modification of level
densities appears quite clearly in \nuc{78}{Ni} also, while neutron levels
around $\varepsilon_\mathrm{F}$ in \nuc{156}{Sn} are shifted in a slightly more
disordered way.

\begin{figure}[htbp]
  \centering
  \includegraphics[width=.49\columnwidth,clip]{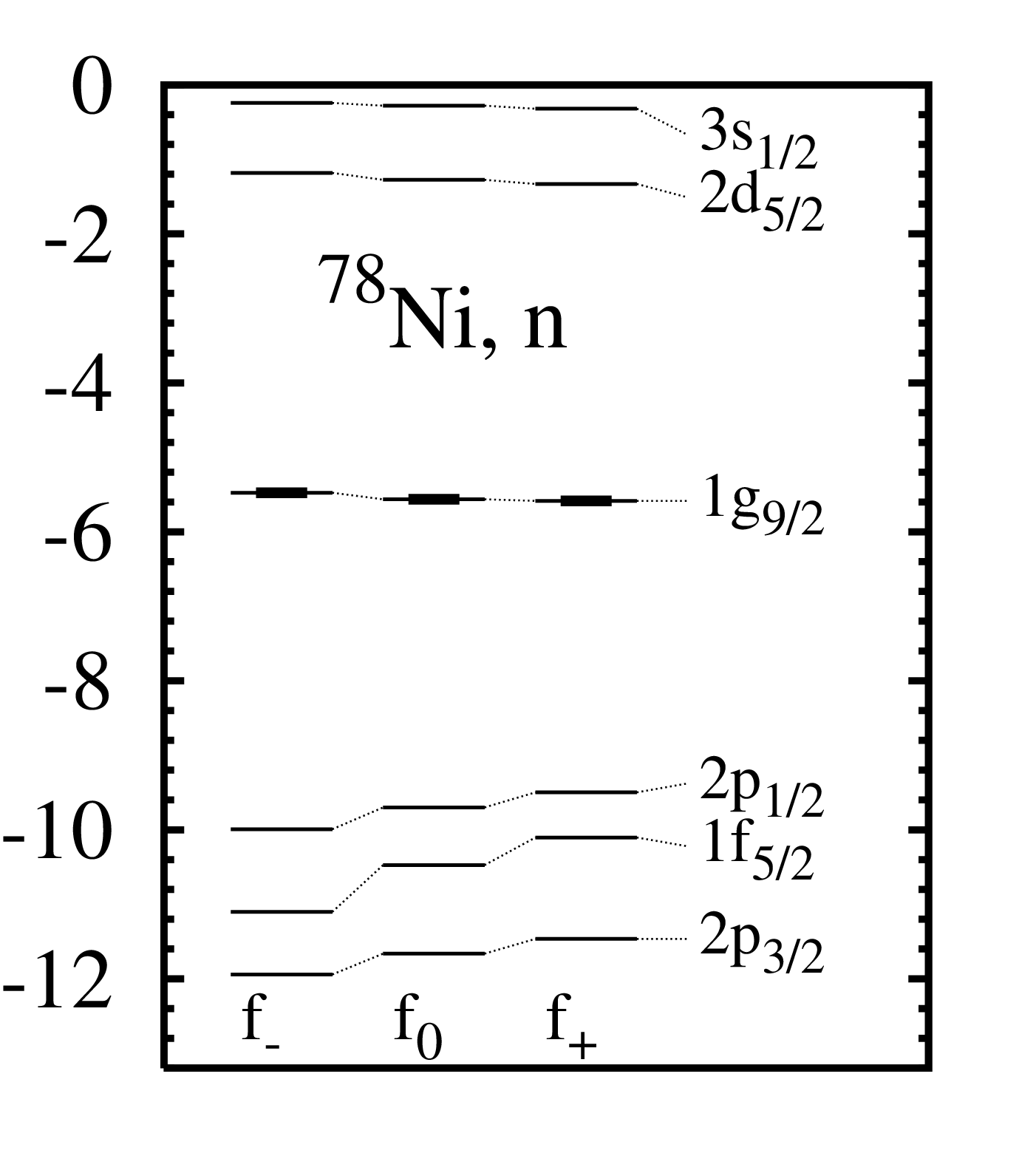}
  \includegraphics[width=.49\columnwidth,clip]{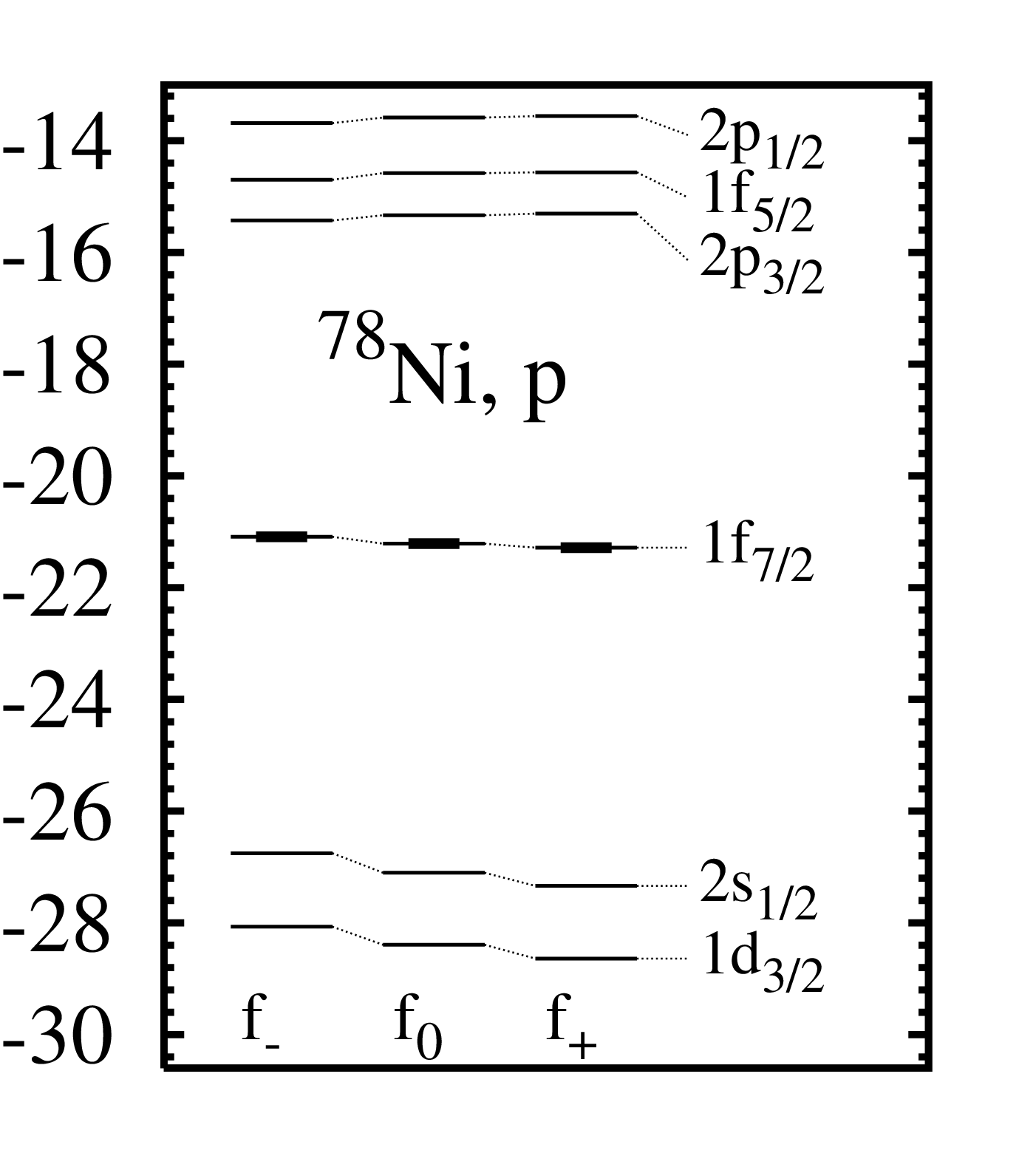} \\
  \includegraphics[width=.49\columnwidth,clip]{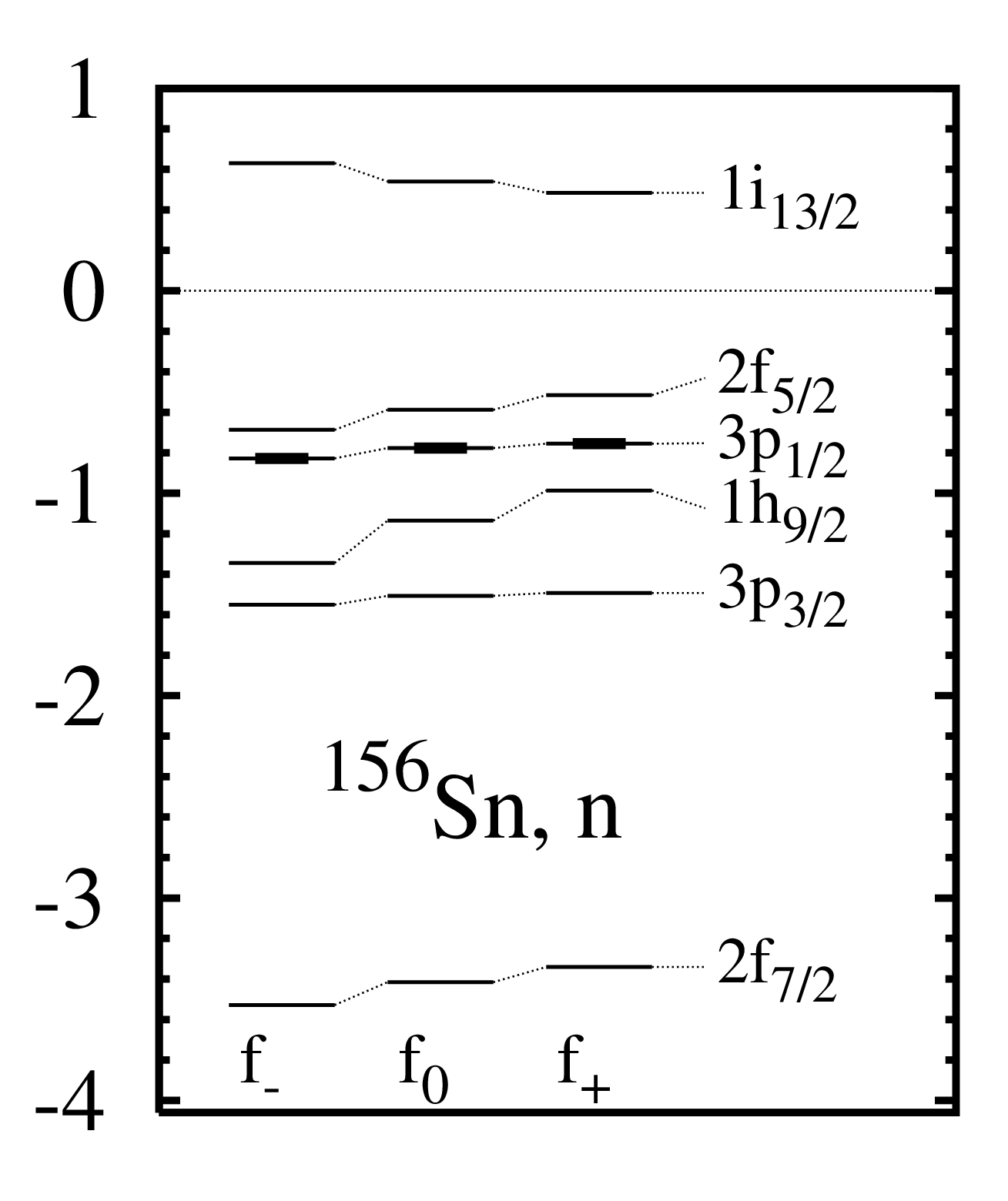}
  \includegraphics[width=.49\columnwidth,clip]{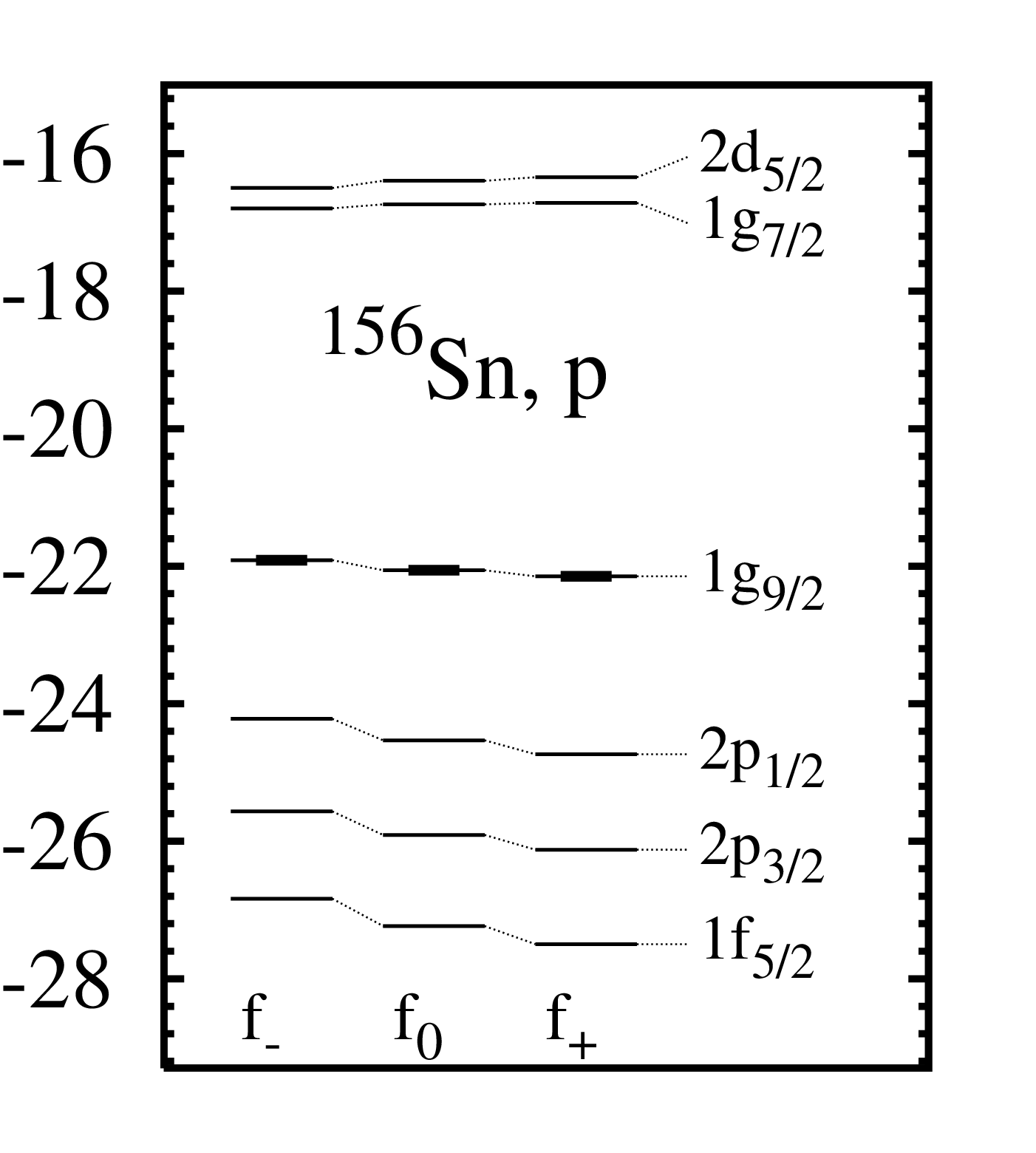} \\
  \caption{Same as \citefigdot{spe1} in \nuc{78}{Ni} and \nuc{156}{Sn}.}
  \label{fig:spe2}
\end{figure}

High-$\ell$/low-$n$ orbitals ($n, \ell$ being respectively the principal
and orbital quantum numbers) are in fact more sensitive to variations of
the spin-orbit field than to $\Delta m^\ast$ because of
their spatial localization near the surface of the nucleus. The
spin-orbit field is modified between functionals by the interplay between
$\Jtens^2$-term coefficients and effective mass parameters, since these both
depend on the same non-local terms of the Skyrme force~\cite{dobaczewski06}.
The spin-orbit force
($\rho \gradvec\cdot\Jvec$ terms in the EDF),
which is subject to a slight readjustment, does affect the spectra as well. We
observed, overall, a marginal increase of the spin-orbit field strength when
going from $f_-$ to $f_+$. This implies that while the global effect of
modifying the level density is quite clearly observed when we alter the
effective mass parameters, details of the spectroscopy are at least as
sensitive to the terms connected to the spin-orbit field.

%---------------------------------------%

\subsubsection{Pairing gaps}

%---------------------------------------%

As an example, neutron spectral gaps are plotted on \citefigdot{gaps} for Sn
and Pb series, up to the drip line, against experimental gaps extracted through
five-point mass formulas~\cite{duguet02a,duguet02b}. The slight change in the
level density translates into a modification of the pairing gaps: a higher
neutron effective mass ($f_+$) corresponds to a denser spectrum and higher
gaps. The effect, which increases with asymmetry, remains however very small,
because of the limited alteration of single-particle levels seen
on~\citefigsdot{spe1}{spe2}.

\begin{figure}[htbp]
  \includegraphics[width=\columnwidth]{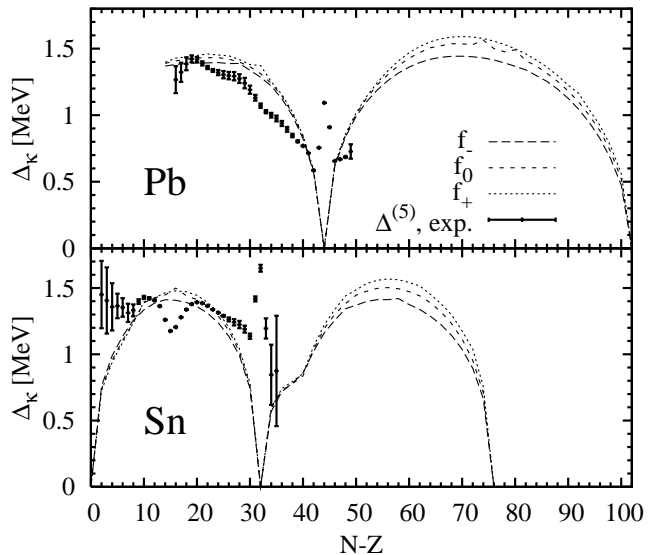}
  \caption{Neutron spectral gaps computed in Sn (bottom) and Pb (top) series
           with $f_-$, $f_0$, $f_+$, as a function of asymmetry. Experimental
           $\Delta^{(5)}$ gaps extracted from masses~\cite{audi2003} are
           plotted with error bars.}
           \label{fig:gaps}
\end{figure}

In the end, the effect is negligible and would be overwhelmed by any
other modification of the particle-hole part of the functional.
For example the spin-orbit force,
acting on the detailed level scheme, could alter the shape of gaps. The
pairing functional itself is a subject of current debate regarding its density
dependence, regularization scheme and finite-range
corrections~\cite{duguetbennaceur}, while the HFB formalism can itself be
improved (particle number projection) as well as the mere choice of observable
to be compared (definition of theoretical an experimental gaps), although our
choice has been proven to be possibly the most sound for extracting pure
pairing effects~\cite{duguet02b}.

%-----------------------------------------------------------------------------%

\subsubsection{Binding energies}

%-----------------------------------------------------------------------------%

Let us now study the effect of the aforementioned variation of level densities
and pairing gaps on binding energies. On \citefigdot{masses} we show the
binding
energy residuals $E_\mathrm{th}-E_\mathrm{exp}$ for Sn and Pb isotopes and
$N=50$ and $N=82$
isotones. The evolution of $E_\mathrm{th}-E_\mathrm{exp}$ along such series
is usually
plagued by an underbinding of open-shell nuclei with respect to closed-shell
ones which translates into an arch
shape of $E$-residual curves. Although the variation of $\emv$ seems to impact
the arches, again, the effect is negligible compared to the absolute value of
deviations from experiment, except in the $N=82$ series where open-shell nuclei
tend to be more underbound in the case of $f_+$.

%.............................................................................%

\begin{figure}[htbp]
  \includegraphics[width=\columnwidth]{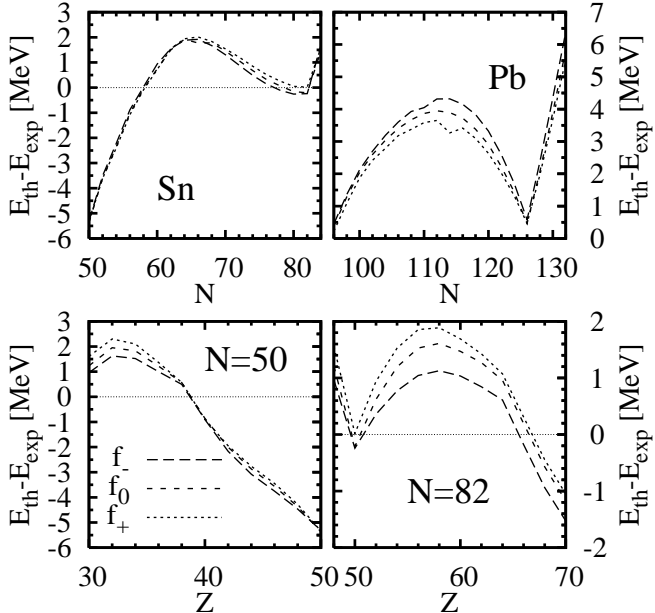}
  \caption{Binding energy residuals computed with forces $f_-$, $f_0$ and
  $f_+$ for semi-magic series of nuclei, as indicated.}
  \label{fig:masses}
\end{figure}

%.............................................................................%

%---------------------------------------%

\subsubsection{Isovector giant resonances}

%---------------------------------------%

The isovector effective mass is usually defined from the energy-weighted
sum rule $\mathfrak{m}_1$ (the TRK sum rule~\cite{trk}) of the isovector giant
dipole resonance (IVGDR):
\begin{equation}
    \mathfrak{m}_1 (E1\;;~T=1) = \frac{\hbar^2}{2m} \frac{NZ}{A}
        \left( 1 + \kapv \right)
        = \frac{\hbar^2}{2m} \frac{NZ}{A} \frac{m}{\emv},
    \label{eq:trk}
\end{equation}
which exhibits its link with the strength distribution of isovector collective
modes. We perform here a schematic study of dynamical properties of $f_-$,
$f_0$, $f_+$ by means of results derived in Ref.~\cite{colo95}. Thanks to RPA
sum rules similar to \citeeqdot{trk}, it is possible to fit an accurate
parameterization of the energy $E_1 = \mathfrak{m}_1/\mathfrak{m}_{-1}$ of
isovector giant resonances in a given nucleus as a function of Skyrme
parameters. Results for GDR ($L=1$) and isovector giant monopole (IVGMR, $L=0$)
modes in \nuc{208}{Pb} are shown in \citetable{ivgr}, compared to experimental
energies (respectively from Refs.~\cite{ritman93} and~\cite{erell86} and
corrected, as suggested in~\cite{colo95}, for the shift due to the spreading of
the strength by damping effects -- 2~MeV for GMR, 1~MeV for GDR).

%.............................................................................%

\begin{table}[htbp]
  \caption{$E_1$ energies of \nuc{208}{Pb} isovector giant resonances computed
  thanks to a sum-rule parameterization (see text), compared to experimental
  energy centroids. Experimental uncertainties are as indicated. We infer
  from figures in Ref.~\cite{colo95} the accuracy of theoretical energies
  computed with the fits in that reference, with respect to full RPA
  calculations, to be of the order of 1~MeV.}
  \smallskip
  \begin{tabular}{lccc}
  \hline
		               & $\kapv$ & $E_1(L=0,T=1)$ & $E_1(L=1,T=1)$ \\
  \hline
		$f_-$          &   0.15  &      24.55     &     12.68      \\
		$f_0$          &   0.43  &      26.43     &     13.60      \\
		$f_+$          &   0.60  &      27.25     &     14.01      \\
  \hline
		exp. centroid  &         & 26.3 $\pm$ 1.1 & 14.3 $\pm$ 0.1 \\
  \hline
  \end{tabular}
  \label{tab:ivgr}
\end{table}

%.............................................................................%

While $f_-$ predicts both energies lower than experimental ones, values
for $f_0$ and $f_+$ are compatible with experiment for the $L=0$ mode, and
only $f_+$ approaches the experimental value for the $L=1$ mode. This suggests
that values of $\kapv$ corresponding to a positive value of $\dmiun$ (equal to
or higher than 0.43 in our case) better describe isovector dynamics than lower
values.

As a summary, the effect of the splitting of neutron and proton effective
masses
with isospin asymmetry on single-particle energies, pairing gaps and binding
energies, is noticeable and consistent, yet limited and thus hardly meaningful
when compared to the overall (in)accuracy of the predictions made by current
nuclear EDF. In fact,
the main reason for not seeing a dramatic modification of EDF predictions when
altering $\dmiun$ is the limited amount of
strongly asymmetric nuclear matter at high enough density in the ground state
of nuclei with realistic isospin as already suggested in~\cite{goriely03}.
This makes the effect of the isovector effective
mass rather marginal. Giant isovector resonances are certainly more fruitful
to seek for an effect of a modification of $\dmiun$. Indeed,
a sum-rule-based analysis of isovector collective modes
allows a slightly more clear-cut conclusion, with a tendency to favor
$\dmiun \gtrsim 0$. The conclusion of the phenomenological study done in this
section is that, while no observable listed here
strongly ask for $\dmiun>0$, there is no reason to omit this constraint in
future functionals, since, as already stated,
ab-initio predictions for the sign of $\dmiun$ are solid.
There remains to check the intrinsic consistency of the functional in terms of
other ab-initio inputs and stability criteria.

%%%%%%%%%%%%%%%%%%%%%%%%%%%%%%%%%%%%%%%%%%%%%%%%%%%%%%%%%%%%%%%%%%%%%%%%%%%%%%%

\section{Further study of infinite matter}
\label{sec:moreinm}

%%%%%%%%%%%%%%%%%%%%%%%%%%%%%%%%%%%%%%%%%%%%%%%%%%%%%%%%%%%%%%%%%%%%%%%%%%%%%%%

\subsection{Separation of the EOS into $(S,T)$ channels}
\label{sec:canauxST}

%=============================================================================%

\subsubsection{Introduction}

%-----------------------------------------------------------------------------%

In this section, we discuss the contributions to the potential energy of SNM
from the four two-body spin-isospin $(S,T)$ channels. We compare our results
with those predicted by BHF calculations~\cite{baldoperso} using
AV18~\cite{argonne} two-body interaction and a three-body force constructed
from meson exchange theory~\cite{grange89a,lejeune00}.

Using projectors on spin singlet and triplet states, respectively
\begin{eqnarray}
    \hat{P}_{S=0} ~=~ \frac{1}{2}( 1 - \Psig ), &~~~&
    \hat{P}_{S=1} ~=~ \frac{1}{2}( 1 + \Psig ),
\end{eqnarray}
where $\Psig$ is the spin-exchange operator, and similar expressions for
isospin projectors $\PT$ using the isospin exchange operator $\Ptau$, yields
the potential energy in each $(S,T)$ channel
\begin{eqnarray}
    E^{ST}_\pot &=&
		\frac{1}{2} \sum_{kl}
		\matel{kl}{V \PS\PT}{\overline{kl}} \rho_{kk} \rho_{ll},
	\label{eq:enerst}
\end{eqnarray}
where the sum on $k,l$ runs over all HF single-particle eigenstates whereas
$\rho_{kk}$ designates the diagonal one-body density matrix. The notation
$\vert \overline{kl} \rangle$ denotes a non-normalized but antisymmetrized
two-body state. In order to compare different
many-body approaches (ab-initio or EDF), we use the ``potential
energy'' which refers to the total binding energy from which is subtracted the
kinetic energy of the non-interacting particle system.

Note that
due to the zero-range character of the Skyrme force, together with at most
second-order derivative terms, only $L=0,1$ partial waves occur explicitly
whereas higher partial waves contribute to ab-initio EOS.
We find, for SNM,
\begin{widesubeqnarray}
  \frac{E^{00}_\pot}{A}
	&=& \frac{3}{160} t_2 (1-x_2)\tauexp,
	\label{eq:inmdecompstart} \\
  \frac{E^{10}_\pot}{A}
	&=& \frac{3}{16} \sum_{i=0}^{2} t_{0i}(1+x_{0i}) \rhoisos^{1+i/3}
		+ \frac{9}{160} t_1 (1+x_1) \tauexp, \\
  \frac{E^{01}_\pot}{A}
	&=& \frac{3}{16} \sum_{i=0}^{2} t_{0i}(1-x_{0i}) \rhoisos^{1+i/3}
		+ \frac{9}{160} t_1(1-x_1) \tauexp, \\
  \frac{E^{11}_\pot}{A} &=&  \frac{27}{160} t_2 (1+x_2) \tauexp\,,
	\label{eq:inmdecompend}
\end{widesubeqnarray}
where $(t_i,x_i)$ are usual coefficients of the Skyrme EDF, whereas
$(t_{0i},x_{0i})$ characterize the density-independent zero-range term and the
two density-dependent ones, following the parameterization of
equation~(\ref{eq:skyrme}).

The coefficients occurring in
Eqs.~(\ref{eq:inmdecompstart})--(\ref{eq:inmdecompend}) stem from the
antisymmetrization condition $(-)^{L+S+T} = -1$, the relative angular momentum
$L$ being even for $t_{0i}$ and $t_1$ ($\kvec^2$) terms and odd for $t_2$
($\kvec'\cdot\kvec$) terms. The expression of the potential energy in channels
$(S, T) = (0, 0)$ and $(1, 1)$ is very simple since only the $t_2$
term contributes.

%-----------------------------------------------------------------------------%

\subsubsection{Force \emph{vs.~}functional}

%-----------------------------------------------------------------------------%

Previous statements, however, apply only to the case where the EDF is computed
as the expectation value of an (antisymmetrized) effective force. In the more
general case, it is still possible to define $(S,T)$ channels starting
from any Hartree-like functional. Indeed, the functional can
always be expressed in terms of an effective non-antisymmetrized vertex and one
can still plug a projector in the calculation of its matrix elements.
In the pure functional case, there is however no more clear definition of
partial waves, and spin-isospin channels emerge from the balance between
coefficients of (iso)scalar/(iso)vector couplings (see \citeappendix{stchan}
for a more detailed discussion).

As long as there are not enough inputs to constrain \emph{all}
degrees of freedom of a general functional, the ``force'' approach remains
as an acceptable path, and hence shall be used in the following.

%-----------------------------------------------------------------------------%

\subsubsection{Results}

Results are plotted against BHF predictions on~\citefigdot{subsp}. First, one
can observe that results are rather scattered. Second, the main source of
binding, from $(S,T)=(0,1)$ and $(1,0)$ channels, is not well described and
the detailed saturation mechanism is not captured. It is clear that, even
though
all four forces reproduce perfectly PNM and SNM EOS, they do not have the
same spin-isospin content, and that the latter is in general rather poor. Thus,
fitting the global EOS is an important element but it does not mean that
spin-isospin properties of the functional are fixed once and for all. One needs
to do more and fitting ab-initio predictions of $E_\mathrm{pot}^{(S,T)}$
seems to be a good idea in the near future. However, one needs to make
sure that the theoretical uncertainty of the data used is smaller than the
expected accuracy of the fit to them. This calls for predictions from other
ab-initio methods using the same two-body plus three-body Hamiltonian.
Then, those ab-initio calculations should be repeated using different sets of
two-body plus three-body Hamiltonians in order to provide a theoretical error
bar on those predictions.

%.............................................................................%

\begin{figure}[htbp]
\centering
    \includegraphics[width=\columnwidth]{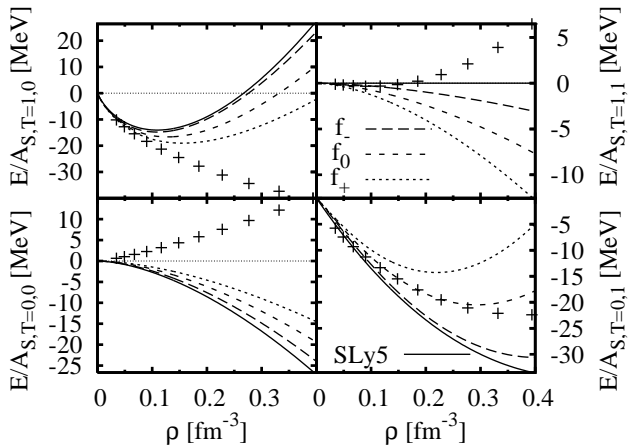}
    \caption{Energy per particle in each $(S,T)$ channel for SNM, as a function
        of density. Crosses refer to the BHF calculations~\cite{baldoperso}.}
    \label{fig:subsp}
\end{figure}

%.............................................................................%

The most obvious discrepancy appears in $(0,0)$ and $(1,1)$ channels where
Skyrme and BHF data have opposite signs above saturation density. The SLy5
parameter set shows a particular behavior in channel $(1,1)$ due to the
choice of $x_2=-1$ to prevent ferromagnetic instabilities in PNM. Note that
these two channels are taken care of, in the Skyrme functional, by the
density-independent $P$-wave term only. The upper-right panel of
\citefigdot{subsp} points out the tendency of Skyrme parameterizations to be
attractive in polarized PNM, and hence to cause a collapse of its EOS at high
density. At lower densities, BHF data show a distinctive behavior, being
slightly attractive below $\rhosat$ and repulsive above. This feature cannot be
matched by the standard Skyrme functional which exhibits a monotonous behavior
as a function of density in this channel, regardless of the value of
$(t_2,x_2)$.

It is also worth noticing that the failure in channel $(1,1)$ becomes more
and more prominent as one makes $\dmiun$ closer to the ab-initio predictions
(force $f_+$). The effective masses being governed by the momentum dependent
terms of the force, it is not a surprise that the modification of the former
impacts channels $(0,0)$ and $(1,1)$. What changes in the coefficients
entering Eqs.~(\ref{eq:inmdecompstart}-\ref{eq:inmdecompend}) stems only from
the variation of $\emv$ and the associated rearrangement of parameters in the
functional, most notably the $\cdrho_{0,1}$ coefficients closely related to
surface and surface-symmetry energies. The relatively tight requirements on the
latter imply that the four parameters of the non-local terms in the standard
Skyrme energy functional would be dramatically overconstrained if we were to
add the $(S,T)$-channel decomposition in the fitting data.

In the end, the rather poor properties of the functional in channels $(0,0)$
and $(1,1)$, the degradation of the latter as the effective mass splitting
is improved, the idea of using ab-initio $(S,T)$ contributions in the fit,
call, at least, for a refinement of the odd-$L$ term in the sense either of a
density dependence or of a higher-order derivative term. The latter being prone
to numerical instabilities and interpretation problems, a density-dependent
$\kvec'\cdot\kvec$ term remains as one of the next potential enhancements to be
brought to the Skyrme EDF (density-dependent derivative terms have
been considered already, but with a focus on even-$L$
terms of the form $t_4(\kvec^2 + \kvec'^2)\rhoisos^\beta$ \cite{pearson97}).
Phenomenological constraints on gradient terms are mainly related to the
surface
of nuclei, i.e. low-density regions. One can expect that, to first order, BHF
data in channel $(S,T)=(1,1)$ can be matched with an extended functional while
retaining a good agreement with other (experimental) data. It is less clear in
channel $(0,0)$ but further exploration of the extended parameter space may
bring Skyrme and BHF data in better agreement.

%=============================================================================%

\subsection{RPA linear response functions and the diagnosis of instabilities}
\label{sec:rpalinres}

%=============================================================================%

We attempt here to study general stability conditions of SNM with respect to
finite-size density, spin, isospin and spin-isospin perturbations. Our basic
ingredient is the RPA response function \cite{fetter71a} derived analytically
in Ref.~\cite{garciarecio92a} for the central part of the Skyrme interaction.
Recent work was done to incorporate the effect of the spin-orbit part which was
found to be quite negligible~\cite{margueron06a}, and will be omitted in the
present work. One starts by defining a one-body perturbing operator
\begin{eqnarray}
{\cal Q}^\chan &=& e^{-i\omega t}~\sum_a e^{i\qvec\cdot\rvec_a}~\Theta_a^\chan,
\label{eq:pert-op}
\end{eqnarray}
where $a$ indexes particles in the system. The one-body spin-isospin
operators $\Theta_a^\chan$ are defined as
\begin{equation}
    \Theta_a^\iss = 1_a,~~~\Theta_a^\ivs = \spin_a,~~~
    \Theta_a^\isv = \ispin_a,~~~ \Theta_a^\ivv = \spin_a\ispin_a,
    \label{eq:st-ops}
\end{equation}
where we use the denomination of (iso-)scalar (s) and (iso-)vector (v) channels
in order to distinguish the uncoupled spin-isospin channels from the coupled
two-body $(S,T)$ channels discussed in the previous section. In
\citeeqdot{st-ops} and the following, the first (second) subscripts
denotes the spin (isospin). We then study the response to each type of
perturbation separately through the \emph{response functions}
\begin{eqnarray}
    \chi^\chan (\omega, \qvec) &=& \frac{1}{\Omega} \sum_n
        \vert\langle n \vert {\cal Q}^\chan \vert 0 \rangle\vert^2
        \nonumber \\ &\times&
        \left(
            \frac{1}{\omega - E_{n0} + i\eta}
            - \frac{1}{\omega + E_{n0} - i\eta}
        \right), ~~~
    \label{eq:respfunc}
\end{eqnarray}
at the RPA level, where $\Omega$ stands for a normalization volume and $\vert
n\rangle$ is an excited state of the system, $E_{n0}$ being the corresponding
energy. Since the central residual interaction does not couple the channels
defined through \citeeqdot{st-ops} in SNM, we can indeed consider each channel
separately.

The response function $\chi^\chan$ can be seen as the propagator of the
collective perturbation, \emph{i.e.~}the positions of its poles in the
$(q, \omega)$
plane yield the dispersion relation of the mode. In this formalism, the onset
of an unstable mode is marked by the occurrence of a pole in $\chi^\chan$ at
$\omega=0$, corresponding to zero excitation energy. Such a pole marks the
transition between stable ($\chi^\chan < 0$) and unstable ($\chi^\chan>0$)
domains. Unstable modes of infinite wavelength ($q=0$) are those traditionally
discussed in the context of Landau parameters. A pole at finite $q$
characterizes a system which is unstable with respect to the appearance of a
spatial oscillation of a given type (density, spin, isospin or spin-isospin)
with a given wavelength $\lambda = 2\pi / q$.

The evaluation of response functions calls for the residual interaction
$V_\ph$,
defined as the second-order functional derivative of the energy with respect to
the density matrix. Its momentum-space matrix elements can be written, using
total momentum conservation, as \cite{garciarecio92a}:
\begin{eqnarray}
    V_\ph(\qvec_1, \qvec_2, \qvec)
        &=& \langle \qvec_1~ \qvec_2+\qvec
            \vert~ V_\ph ~
            \vert \qvec_1+\qvec~ \qvec_2 \rangle,~~~~ \label{eq:vphkspace} \\
        &=& \hat{W}_1(q) + \hat{W}_2(q)~(\qvec_1 - \qvec_2)^2,
\end{eqnarray}
with
\begin{eqnarray}
    \hat{W}_1(q) &=& \frac{1}{4}~[~W_1^\iss(q) + W_1^\ivs(q)~\sigud
        + W_1^\isv(q)~\tauud  \nonumber \\
        && +~ W_1^\ivv(q)~\sigud~\tauud~],
\end{eqnarray}
and a similar expression for $\hat{W}_2$. We find:
\begin{widetext}
    \raggedright
    \begin{tabular}{p{12cm}p{5cm}}
    \begin{subeqnarray}
        \label{eq:wfuncsdefbegin}
        \frac{W_1^\iss(q)}{4} &=& 2 \crhoz_0
            + \sum_{i=1}^2 \crhoi_0 
\frac{(i+6)(i+3)}{9}
 \rhoit
            - 2 \cdrho_0 \qvec^2, \\
        \frac{W_1^\ivs(q)}{4} &=& 2 \csz_0
            + 2 \sum_{i=1}^2 \csi_0 \rhoit - 2 \cds_0 \qvec^2,
            \\
        \frac{W_1^\isv(q)}{4}
            &=& 2 \crhoz_1
            + 2 \sum_{i=1}^2 \crhoi_1 \rhoit - 2 \cdrho_1 \qvec^2, \\
        \frac{W_1^\ivv(q)}{4} &=& 2 \csz_1
            + 2 \sum_{i=1}^2 \csi_1 \rhoit - 2 \cds_1 \qvec^2,
    \end{subeqnarray} &
    \begin{subeqnarray}
        \raisebox{0pt}[1.7em][1.5em]{$\displaystyle\frac{W_2^\iss(q)}{4}$}
			&=& \ctau_0, \\
        \raisebox{0pt}[1.6em][1.5em]{$\displaystyle\frac{W_2^\ivs(q)}{4}$}
			&=& \csT_0,  \\
        \raisebox{0pt}[1.6em][1.5em]{$\displaystyle\frac{W_2^\isv(q)}{4}$}
			&=& \ctau_1,  \\
        \raisebox{0pt}[1.6em][1.5em]{$\displaystyle\frac{W_2^\ivv(q)}{4}$}
			&=& \csT_1.
        \label{eq:wfuncsdefend}
    \end{subeqnarray}
    \end{tabular}
	With the above expression for the residual interaction, the response
	function reads as
	\begin{eqnarray}
		\chi^\chan &(\omega, \qvec)& = 4\Pi_0
		\left[\raisebox{0pt}[2em][1em]{~}\right.
		1 - W_1^\chan\Pi_0
	        - 2 W_2^\chan \kfermi^2 \left(
		\qbar^2  - \frac{\nu^2}{1 - \frac{m^\ast \kfermi^3}{3\pi^2}~
					W_2^\chan}
		\right) \Pi_0  \\
		&+& 2 W_2^\chan \kfermi^2 (2\qbar^2~ \Pi_0 - \Pi_2)
		 + (W_2^\chan \kfermi^2)^2 \left(
			\Pi_2^2 - \Pi_0 \Pi_4 + 4\qbar^2\nu^2\Pi_0^2
			- \frac{2 m^\ast \kfermi}{3\pi^2} \qbar^2 \Pi_0
		\right)
		\left.\raisebox{0pt}[2em][1em]{~}\right] ^{-1},\nonumber
	\end{eqnarray}
\end{widetext}
where $\qbar = q/2\kfermi$, $\nu = \omega \ems / q \kfermi$ and $\Pi_{0,2,4}$
are generalized Lindhard functions, see Ref.~\cite{garciarecio92a}.

As already said, the limit $\qvec \rightarrow 0$ corresponds to perturbations
of infinite wavelength, keeping the system homogeneous. There, the residual
interaction is uniquely determined by Landau parameters $F_l, F'_l, G_l, G'_l$,
with $l = 0, 1$, and well known stability conditions are obtained under the
form~\cite{migdal67a}:
\begin{eqnarray}
    1 + \frac{X_l}{2l+1} > 0,
    \label{eq:landaustab}
\end{eqnarray}
where $X_l$ represents any of the Landau parameters. We have used this
criterion in the fit of our forces $f_x$, ensuring that no spin or spin-isospin
instability would occur below $2\rhosat$. We observe that, from the point of
view of Landau parameters, the most critical channel is the vector-isovector
one, with associated instabilities at densities as low as $2\rho_{sat}$
(see the upper-right panel of \citefigdot{response}). This behavior is linked
to the attractive
character of the functional in channel $(S,T) = (1,1)$ which gives rise to
a collapse of spin-polarized PNM, and accordingly, a vanishing spin-isospin
symmetry energy. Therefore, better reproducing the decomposition into $(S,T)$
channels of EOS obtained from ab-initio methods is not only a matter of
microscopic motivation, but also a necessity to avoid unwanted instabilities.

Beyond infinite wavelength instabilities, we also aim at demonstrating that a
more general treatment is needed to fully describe and control unstable modes
which arise in the Skyrme EDF framework. Thus, contributions to the residual
interaction coming from functional terms of the form $\rho\Delta\rho$ are zero
for $\qvec=0$, whereas such terms drive finite-size instabilities.

Indeed, we have observed that existing (SkP) or new parameterizations built
with a high value of $\kappa_v$ in order to reproduce the microscopic splitting
of effective masses, tend to spatially separate protons from neutrons in
spherical mean-field calculations, where enough iterations lead to states with
strongly oscillating densities and a diverging energy. Following a preliminary
phenomenological reasoning, we could relate this effect to the
$\cdrho_1 \rhoisov\cdot\Delta\rhoisov$ term in the functional, as this term
can energetically favor strong oscillations of the isovector density $\rhoisov$
which arise in the case of such a spatial n-p separation. Moreover,
Eqs.~(\ref{eq:wfuncsdefbegin}-\ref{eq:wfuncsdefend}) show that such a term
can yield an attractive contribution to the residual interaction in the case of
a short-wavelength (high $q$) perturbation. We found empirically that parameter
sets for which this instability arises are characterized by a high value of
$\cdrho_1$, that is $\cdrho_1 \gtrsim 30$. As seen from \citetable{cun},
this parameter is strongly correlated with the effective mass splitting
$\dmiun$ in such a way that a positive splitting, as required by ab-initio
predictions, leads to instabilities.

%.............................................................................%

\begin{table}
    \caption{Values of the $\cun$ coefficient, in MeV~fm$^{5}$.}
    \label{tab:cun}
    \begin{center}
    \begin{tabular}{ccccccc}
        \noalign{\smallskip}\hline\noalign{\smallskip}
                  & $f_{-}$ & SLy5    & $f_{0}$ & $f_{+}$ & LNS   & SkP   \\
        \noalign{\smallskip}\hline\noalign{\smallskip}
        $\dmiun$  & -0.284  & -0.182  & 0.001   & 0.170   & 0.227 & 0.418 \\
        \noalign{\smallskip}\hline\noalign{\smallskip}
        $\cun$    & 5.4     & 16.7    & 21.4    & 29.4    & 33.75 & 35.0  \\
        \noalign{\smallskip}\hline
    \end{tabular}
    \end{center}
\end{table}

%.............................................................................%

Whereas we were obviously unable to provide a fully converged (and hence
physically meaningful) and clearly unstable force to illustrate the previous
statements, we found that certain forces available in the literature present
the aforementioned behavior. For example, convergence problems
have arisen (and have already been pointed out in another study~\cite{engel06})
for the SkP parameter set~\cite{doba84a}. The nature of the instabilities
discussed here is illustrated on the lower panels of \citefigdot{skp-ni56},
where neutron and proton densities are plotted at various stages of execution
of a self-consistent iterative procedure with SkP in \nuc{56}{Ni}. We see that
strong, opposing oscillations of neutron and proton densities are formed, and
steadily increase with iterations. Such a behavior happens after a seemingly
converged situation for which the relative energy variation is small but almost
constant over a large number of iterations and the evolution of the energy
is monotonous.

%.............................................................................%

\begin{figure}[htbp]
    \includegraphics[width=.98\columnwidth]{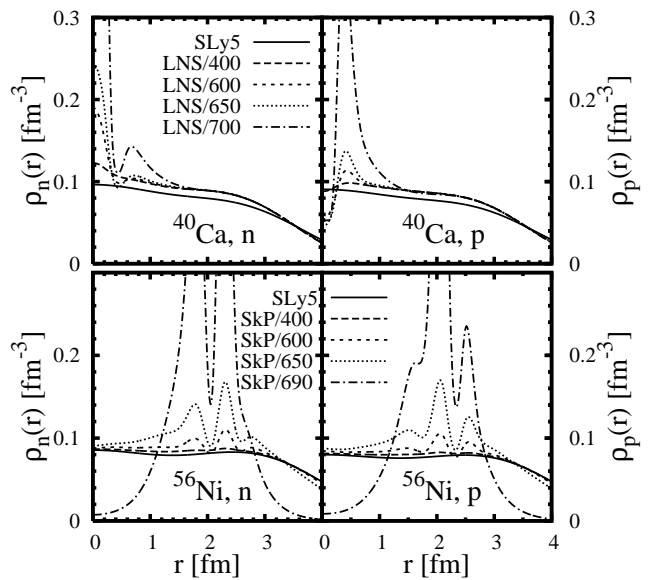}
    \caption{Neutron and proton densities in central regions of
  \nuc{56}{Ni} (bottom) and \nuc{40}{Ca} (top) plotted for a fully
  converged computation using the SLy5 force (solid line; relative variation
  of energy between iterations less than $10^{-14}$) and along a series of
  iterations done with SkP (bottom) and LNS (top).
  Number of iterations as indicated in key.
  In both cases the collapse happens after a seemingly
  converged situation ($\sim 10^{-9}$ relative energy variation, steady
  over a large number of iterations indicating a nearly linear evolution
  of the energy), which can be mistaken for an energy minimum if too loose a 
  convergence criterion is used.}
  \label{fig:skp-ni56}
\end{figure}

%.............................................................................%

The study of the linear response function in the scalar-isovector channel
allows us to provide a more quantitative ground to the previous observation. By
plotting critical densities (lowest density of occurrence of a pole in
$\chi^\chan(\omega=0,q)$) for a given $q$ on \citefigdot{response-tau},
we see that these critical densities can be very close to $\rhosat$ for $q
\approx 2.5$ to $3 \fmu$. This is the case for SkP, which displays the lowest
scalar-isovector critical density of all forces studied in this paper.
Accordingly, it is the most prone to a lack of convergence in HF calculations.

%.............................................................................%

\begin{figure}[htbp]
    \includegraphics[width=.9\columnwidth]{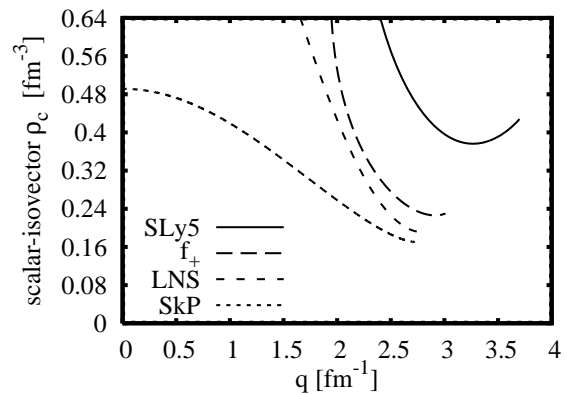}
    \caption{The lowest density of occurrence of a pole in
        \mbox{$\chi^\isv (\omega=0, \qvec)$}
        is plotted against the wave-number
        $q$ of the scalar-isovector perturbation.
The curves end at $q=2k_F$ since the ground state can not couple to
excitations with $\omega = 0, q > 2k_F$.}
    \label{fig:response-tau}
\end{figure}

%.............................................................................%

The link between response functions and convergence problems can indeed be
understood by classifying them by their magnitude: in case of a stable but very
soft mode, lack of convergence arises from the existence of a continuum of
quasi-degenerate mean-field states, among which no minimization or
self-consistency algorithm shall be able to decisively find an energy minimum
without a considerable amount of iterations. If the soft mode becomes unstable,
it causes a divergence of the energy and of other observable such as the
densities. We see in the agreement between the RPA study of SNM and the
observation of unstable HF calculations of nuclei a qualitative validation of
our Local Density Approximation (LDA)-based treatment of instabilities: soft
or unstable modes occurring in INM happen for the same parameter sets in finite
nuclei.

The large number of iterations needed for the divergence to occur on
\citefigdot{skp-ni56} is a consequence of the limiting case embodied by SkP,
such that the existence of a definite instability is highly dependent on finite
size effects (choice of the nucleus) and discretization details in the
numerical procedure. If SkP is a limiting case, LNS also displays a low
critical
density in the scalar-isovector channel (\citefigdot{response-tau}). In this
case, we observed proton-neutron separation in \nuc{40}{Ca} and for small mesh
steps ($0.1$~fm) only (see \citefigdot{skp-ni56}), while it is more frequent
with SkP. Our force $f_+$, with a critical density just slightly higher than
LNS, successfully passed the test of computing a series of 134 spherical
nuclei.
This again demonstrates that testing finite-size instabilities through
response functions constitutes an accurate tool, the critical density (and its
proximity to $\rhosat$) being a good measure of the gravity of the problems
one might encounter in finite nuclei. Although the actual occurrence of
instabilities is subject to details of the numerical treatment, it is now clear
that their origin can be traced back to the choice of parameters in the
functional itself.

Nevertheless, even if a functional does not display clear instabilities but
only spurious soft collective modes, convergence difficulties shall arise in
mean-field calculations while such a mode will translate into a non-physical
low-lying spectrum in a beyond-mean-field framework. This can then
yield excessive correlation energies if one systematically includes
correlations in the ground state \emph{e.g.~}in (Q)RPA or GCM methods. One
should thus make sure that no spurious (even remotely) soft mode occurs at
saturation density in order to prevent such problems.

Having demonstrated the importance of finite-size instabilities, let us go back
to discussing our original set of forces and perform a generalization to other
spin-isospin channels.

Critical densities are plotted on \citefigdot{response} for the four channels
defined in \citeeqdot{st-ops}. The upper-left panel shows that, while no
unstable mode occurs at $q=0$ thanks to fitting PNM EOS to relatively high
density, scalar-isovector instabilities may happen little above $\rhosat$ for
$q\approx2.5$ to $3\fmu$. In addition, there is a clear trend for lowering the
critical density when $\dmiun$ is increased, in agreement with the preliminary
phenomenological reasoning on $\cdrho_1$.

%.............................................................................%

\begin{figure}[htbp]
    \includegraphics[width=\columnwidth]{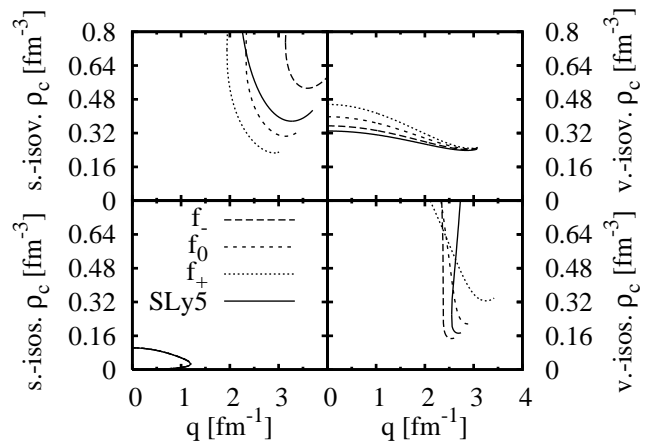}
    \caption{Same as \citefigdot{response-tau}, for all spin-isospin channels.
    The lower-left panel shows the region of spinodal instabilities below
    $\rhosat$. The domain of $q$ covered in this case determines the size of
    structures formed, while the region between $0.1$ and $0.16\fmu$ appears as
    metastable.}
    \label{fig:response}
\end{figure}

%.............................................................................%

Spin channels have been taken care of during the fit thanks to Landau
parameters, which describe the residual interaction at $q=0$. The result can
be seen on the right panels of \citefigdot{response}, where the critical
densities of instability are plotted for spin-flip modes (isoscalar and
isovector). As previously stated, the most dangerous $q=0$ instability is found
in the vector-isovector channel. By looking at the upper-right panel of
\citefigdot{response} one can see that the critical density is even lower
($1.5\rhosat$) at $q=3\fmu$ than at $q=0$, a domain not covered by the
criterion of~\citeeqdot{landaustab}.

An even more prominent finite-size effect can be observed in the isoscalar
spin-flip channel (lower-right panel of \citefigdot{response}) where, while
no instability occurs at $q=0$ as in the case of most Skyrme forces,
finite-size instabilities occur at low density, even below $\rhosat$ !
These instabilities are linked to the $\cds_0~\sisos\cdot\Delta\sisos$ term
which makes the vector-isoscalar $V_\ph$ attractive at large $q$ whereas it is
repulsive at $q=0$. Values of $\cds_0$, indeed, are as high as $45.85$ and
$47.32$ for SLy5 and $f_-$, respectively. As a consequence, one can expect
divergences in calculations of odd or rotating nuclei with the latter forces if
the aforementioned terms are included. In this case, though, increasing
$\dmiun$ pushes the critical density farther from $\rhosat$: $f_0$ and $f_+$
functionals are thus the only ones to be free from instabilities at $\rhosat$,
without being totally satisfactory either.

The previous discussion is valid if the full time-odd functional is
taken into account. This must be stressed since $\sisos\cdot\Delta\sisos$
terms, which drive the most critical, finite-size instabilities, have never
been included in self-consistent mean field calculations to date. However, RPA
calculations are commonly performed by computing the residual interaction
matrices directly from the antisymmetrized force (plus rearrangement terms),
which amounts to implicitly including the contribution to $V_\ph$ from all
terms in the functional~\cite{engel05}.

The latter findings finalize the picture of a competition between spin and
isospin instabilities. All in all, the strong interplay between the
various quantities linked to the four parameters of the non-local terms in
the Skyrme force does not seem to allow for a fully satisfactory compromise
between stability criteria and ab-initio constraints on $\dmiun$.
Again, we see that the non-local part of the Skyrme force is too simplistic to
control all relevant properties. An extension with density- and
momentum-dependent terms, allowing the fine-tuning of the functional at various
densities, combined with the formal checks advocated in this paper, could prove
to significantly improve the predictive power of Skyrme EDF.

%%%%%%%%%%%%%%%%%%%%%%%%%%%%%%%%%%%%%%%%%%%%%%%%%%%%%%%%%%%%%%%%%%%%%%%%%%%%%%%

\section{Conclusion}

%%%%%%%%%%%%%%%%%%%%%%%%%%%%%%%%%%%%%%%%%%%%%%%%%%%%%%%%%%%%%%%%%%%%%%%%%%%%%%%

We have built a series of Skyrme energy density functionals to study the effect
of a variation of the splitting of neutron and proton effective masses with
isospin asymmetry on properties of this mean-field-based model. Thanks to the
use of a second density-dependent term in the underlying effective force, we
could cover a wide range of effective mass splittings ($\dmiun$) with a
satisfactory fit to nuclear properties. Indeed, nuclear observable predicted by
our functionals $f_-$, $f_0$ and $f_+$ show a remarkable similarity, pointing
out that spectra, pairing gaps and masses of bound nuclei are weakly sensitive
to $\dmiun$, mostly due to their relatively low isospin asymmetry. Although
observable were affected in a noticeable and consistent way, no clear
improvement was seen when altering $\dmiun$ either way.

Beyond this phenomenological study, we have compared the splitting of the
equation of state of symmetric infinite matter into spin-isospin channels
provided by our functionals
and by ab-initio Brueckner-Hartree-Fock calculations. Such a comparison showed
an obvious discrepancy in $(S,T)=(0,0)$ and $(1,1)$ channels, where energies
predicted by Skyrme functionals and by BHF calculations have opposite
signs. The inconsistency in channel $(S,T)=(1,1)$, where the Skyrme functional
is attractive, translates into a collapse of polarized neutron matter EOS,
related to the onset of spin-isospin instabilities at quite low density
($2\rhosat$). In this channel, ab-initio predictions cannot be matched (in the
Skyrme ``force'' approach) without an extension of the P-wave term. We also
identified finite-size isospin instabilities caused by strong isovector
gradient terms, which prevent the convergence of mean-field calculations. We
were able to provide a firm and quantitative basis to these observations
through an analysis of finite-size instabilities by use of RPA linear response
functions in SNM. The latter showed that finite-size effects in the analysis
of instabilities tend to always dominate.

The present study leads us to propose the systematic inclusion of consistency
checks with ab-initio predictions of spin-isospin properties in the
construction of our future functionals, as well as a systematic diagnosis of
finite-size instabilities.

%%%%%%%%%%%%%%%%%%%%%%%%%%%%%%%%%%%%%%%%%%%%%%%%%%%%%%%%%%%%%%%%%%%%%%%%%%%%%%%
%%%%%%%%%%%%%%%%%%%%%%%%%%%%%%%%%%%%%%%%%%%%%%%%%%%%%%%%%%%%%%%%%%%%%%%%%%%%%%%

\begin{acknowledgments}
Work by K.~B. was performed within the framework of the Espace de
Structure Nucl\'eaire Th\'eorique (ESNT).
The authors are grateful to D.~Lacroix for a careful reading of the manuscript
and helpful comments and also thank M.~Bender and B.~Cochet for valuable
discussions during the development of this study. Two of us (K.~B. and T.~L.)
wish to thank the NSCL for its hospitality during final stages of this work.
This work was supported by the U.S. National Science Foundation under Grant No.
PHY-0456903.
\end{acknowledgments}

%%%%%%%%%%%%%%%%%%%%%%%%%%%%%%%%%%%%%%%%%%%%%%%%%%%%%%%%%%%%%%%%%%%%%%%%%%%%%%%
%%%%%%%%%%%%%%%%%%%%%%%%%%%%%%%%%%%%%%%%%%%%%%%%%%%%%%%%%%%%%%%%%%%%%%%%%%%%%%%

\appendix

%%%%%%%%%%%%%%%%%%%%%%%%%%%%%%%%%%%%%%%%%%%%%%%%%%%%%%%%%%%%%%%%%%%%%%%%%%%%%%%

\section{Skyrme energy functional}
\label{app:dft}

%%%%%%%%%%%%%%%%%%%%%%%%%%%%%%%%%%%%%%%%%%%%%%%%%%%%%%%%%%%%%%%%%%%%%%%%%%%%%%%

We take the particle-hole part of the functional as given by
the expectation value of a Skyrme effective force including
two density-dependent terms:
\begin{eqnarray}
V(\Rvec,\rvec)
  &=& \sum_{i=0}^2 t_{0i} \left( 1 + x_{0i} \Psig \right) \delr
      \left[ \rhoisos\left(\Rvec\right) \right]^{i/3}
      \nonumber \\
  &+& \demi t_1 \left( 1 + x_1 \Psig \right) \,
      \left[ \delr~ \kvec^2 + \kvec'^2~ \delr \right] \nonumber \\
  &+& t_{2} \left( 1 + x_2 \Psig \right) \kvec' \cdot \delr~ \kvec \nonumber \\
  &+& i W_0 \left[ \sigvec_1 + \sigvec_2 \right]
            \kvec' \times \delr~ \kvec,
    \label{eq:skyrme}
\end{eqnarray}
with the usual notations
\begin{subeqnarray}
    \Rvec   &=& \left( \rvec_1 + \rvec_2 \right)/2, \\
    \rvec   &=& \rvec_1 - \rvec_2, \\
    \kvec   &=& \frac{1}{2i}~\left( \gradvec_1 - \gradvec_2 \right), \\
    \kvec'  &=& {\mbox{C.C. of $\kvec$ acting on the left}},\\
    \sigvec &=& \sigvec_1 + \sigvec_2, \\
    \Psig   &=& \demi~\left( 1 + \sigvec_1 \cdot \sigvec_2 \right).
\end{subeqnarray}

The total binding energy of a nuclear system can
be written as a functional of a local energy density
\begin{eqnarray}
E       &=& \int \mathrm d^3\rvec~ \CH(\rvec), \\
\CH     &=& \frac{\hbar^2}{2m} \tauisos + \CH_\Sk + \CH_\Coul, \\
\CH_\Sk &=& \CH_0^\even + \CH_1^\even + \CH_0^\odd + \CH_1^\odd,
\end{eqnarray}
where the superscripts in the last equation indicate the behavior with
respect to time reversal of densities occurring in each term, while subscripts
indicate the rank of the densities in isospin space. The
corresponding expressions are
\begin{widesubeqnarray}
\CH_0^\even &=&
    \crho_0~\rhoisos^2 + \cdrho_0~\rhoisos \Delta~\rhoisos
    + \ctau_0~\rhoisos~\tauisos
    + \cJ_0~\Jisos^2 + \cdJ_0~\rhoisos \gradvec\cdot\Jvisos,
                                 \label{eq:skfunc1}\\
\CH_1^\even &=&
    \crho_1~\rhoisov^{~2} + \cdrho_1~\rhoisov \cdot \Delta \rhoisov
    + \ctau_1~\rhoisov \cdot \tauisov
    + \cJ_1~\Jisov^{~2} + \cdJ_1~\rhoisov \cdot \gradvec\cdot\Jvisov, \\
\CH_0^\odd &=&
    \cs_0~\sisos^2 + \cds_0~\sisos\cdot\Delta\sisos
    + \csT_0~\sisos\cdot\Tisos + \cgs_0~(\gradvec\cdot\sisos)^2+\cj_0~\jisos^2
    + \cdj_0~\sisos\cdot(\gradvec\times\jisos), \\
\CH_1^\odd &=&
    \cs_1~\sisov^{~2} + \cds_1~\sisov\cdot\Delta\sisov
    + \csT_1~\sisov\cdot\Tisov + \cgs_1~(\gradvec\cdot\sisov)^2
    + \cj_1~\jisov^{~2} + \cdj_1~\sisov\cdot(\gradvec\times\jisov).
\label{eq:skfunc4}
\end{widesubeqnarray}
Bold letters denote vector densities and
arrows denote isovector densities. Neutron and proton densities ($q = +1$ for
neutrons, $q = -1$ for protons) are thus given by
\begin{eqnarray}
    \rho_q \left( \rvec \right)
        &=& \demi\left( \rhoisos + q~ \rho_{1,3} \right), \nonumber \\
    \tau_q \left( \rvec \right)
        &=& \demi\left( \tauisos + q~ \tau_{1,3} \right), \nonumber \\
    \Jtens_q \left( \rvec \right)
        &=& \demi\left( \Jtens_0 + q~ \Jtens_{1,3} \right), \label{eq:rho}
\end{eqnarray}
with similar expressions for time-odd densities. The spin-current
vector $\Jvec_t$ is built from the antisymmetric part of tensor $\Jtens_t$.

Let us give the expressions, in terms of Skyrme force parameters, of the
coupling constants entering the HFB calculations which are
altered by the addition of a second density-dependent term
\begin{subeqnarray}
	\crho_0 &=& \sum_{i=0}^2~\crhoi_0~\rhoisos^{i/3}
		~=~ \sum_{i=0}^2~\thuit~t_{0i}~\rhoisos^{i/3},\\
	\crho_1 &=&  \sum_{i=0}^2~\crhoi_1~\rhoisos^{i/3} \nonumber \\
		&=& \sum_{i=0}^2~-\huit~t_{0i}~(1+2x_{0i})~\rhoisos^{i/3},
\end{subeqnarray}
as well as the constants related, respectively, to isoscalar and isovector
effective masses,
\begin{subeqnarray}
	\ctau_0 &=& \seize \left[~ 3 t_1 + t_2 ( 5 + 4x_2 ) ~\right],\\
	\ctau_1 &=& \seize \left[~ - t_1 ( 1 - 2x_1 ) + t_2 (1+2x_2) ~\right],
\end{subeqnarray}
and constants multiplying gradient terms discussed in \citesecdot{rpalinres},
\begin{subeqnarray}
	\cdrho_0 &=& \squat \left[~ -9 t_1 + t_2 ( 5 + 4x_2 ) ~\right],\\
	\cdrho_1 &=& \squat \left[~ 3t_1 ( 1 + 2x_1 ) + t_2 (1+2x_2)~\right],\\
	\cds_0 &=& \squat \left[~ 3t_1 ( 1 - 2x_1 ) + t_2 (1+2x_2)~\right],\\
	\cds_1 &=& \squat \left[~ 3t_1 + t_2 ~\right].
\end{subeqnarray}
The expressions of all other coupling constants are given in
Ref.~\cite{bender03b}.
Some of the above constants are linked through the gauge invariance of the
functional, related to the Galilean invariance of the underlying effective
force:
\begin{equation}
    \cj_t = -\ctau_t,~~~\cJ_t = -\csT_t,~~~\cdj_t = \cdJ_t.
\end{equation}

The densities used above can be written as functionals of the density matrix
expressed in coordinate space, \emph{i.e.~}
\begin{eqnarray}
	\rhomat &=& \sum_k \langle k \vert \xvec' \sigma' q' \rangle
		\langle \xvec \sigma q \vert k \rangle~ \rho_{kk},
\end{eqnarray}
as
\begin{subeqnarray}
  \rhoisos(\rvec) &=& \sumrsq \deltarx \deltax~ \deltaq \deltas \nonumber \\
  &\times& \rhomat, \\
  \label{eq:densdefbegin}
  \Delta\rhoisos(\rvec) &=& \sumrsq \deltarx \deltax~ \deltaq \deltas
    \nonumber \\
    &\times& (\gradvec'^2 \!+ 2\gradvec'\cdot\gradvec +
    \gradvec^2)\,
    \rhomat, \\
  \tauisos(\rvec) &=& \sumrsq \deltarx \deltax~ \deltaq \deltas \nonumber \\
    &\times& \gradvec'\cdot\gradvec~ \rhomat, \\
  \Jisos(\rvec) &=& \sumrsq \deltarx \deltax~ \deltaq \nonumber \\
    &\times& \frac{1}{2i} (\gradvec-\gradvec')\otimes\spinmat~ \rhomat,\\
  \jisos(\rvec) &=& \sumrsq \deltarx \deltax~ \deltaq \deltas \nonumber \\
    &\times& \frac{1}{2i} (\gradvec - \gradvec')~ \rhomat,
  \label{eq:densdefend}
\end{subeqnarray}
where $\sigma$, $\sigma'$ are indices referring to spin, $q$ and $q'$
refer to isospin, $\gradvec$ is the gradient operator acting on
the coordinate $x$, $\gradvec'$ being the same acting on $x'$. Isovector and
other time-odd densities can be expressed by replacing, respectively,
$\deltaq$ by $\ispinmat$ and $\deltas$ by $\spinmat$ where appropriate.

%%%%%%%%%%%%%%%%%%%%%%%%%%%%%%%%%%%%%%%%%%%%%%%%%%%%%%%%%%%%%%%%%%%%%%%%%%%%%%%

\section{Separation of the energy into spin-isospin channels}
\label{app:stchan}

%%%%%%%%%%%%%%%%%%%%%%%%%%%%%%%%%%%%%%%%%%%%%%%%%%%%%%%%%%%%%%%%%%%%%%%%%%%%%%%

When the EDF is defined as the expectation value of an
effective Hamiltonian, separating it into spin-isospin channels is
straightforward, as in \citeeqdot{enerst}. However, one can extend this
definition to the case of any Hartree-like functional: let us start by
recalling that in the case of the Skyrme force, the direct and exchange terms
have the same analytical structure; one thus usually uses the expressions
\begin{eqnarray}
	E_\pot &=& \frac{1}{2}
           \sum_{kl} \matel{kl}{V_\Sk}{\overline{kl}}~ \rho_{kk}~\rho_{ll},
		\\
	\ket{\overline{kl}} &=& \ket{kl} - \ket{lk}
		~=~ (1-\hat{P}_r\Psig\Ptau)\ket{kl},
\end{eqnarray}
where the last expression uses the position, spin and isospin exchange
operators to define an antisymmetrized and non-normalized two-body state. One
then writes down the antisymmetrized form of \citeeqdot{skyrme} and the EDF by
using the definition of densities entering
Eqs.~(\ref{eq:skfunc1})-(\ref{eq:skfunc4}).

Leaving the antisymmetrized Hamiltonian framework, it is always possible to
define the potential part of the functional as the \emph{direct term} of the
expectation value of a certain operator, as in
\begin{eqnarray}
	E_\pot &=& \sum_{kl} \matel{kl}{V_\EDF}{kl}~ \rho_{kk}~\rho_{ll},
\end{eqnarray}
recalling that $V_\EDF = V_\Sk(1-\hat{P}_r\Psig\Ptau)$ in the Hamiltonian case.
One then defines the energy per channel as
\begin{eqnarray}
	E_\EDF^{ST} &=& \sum_{kl} \matel{kl}{V_\EDF~\PS\PT}{kl}~
		\rho_{kk}~\rho_{ll},
\end{eqnarray}
which, with the definitions (\ref{eq:skfunc1})-(\ref{eq:skfunc4}) for coupling
constants, yields (retaining only terms acting in infinite matter)
\begin{wideeqnarray}
	E_\pot^{ST} &=& \int \mathrm d^3\rvec~\CH^{ST}(\mathbf r) \\
	\CH^{ST} &=&
	\big[ \crho_0 + (4S-3) \cs_0 + (4T-3) \crho_1
		+ (4S-3)(4T-3)\cs_1 \big] \nonumber \\ &&
	~\times~ \frac{1}{16}~ \big[
		(2S+1)(2T+1)\rhoisos^2 + (2S-1)(2T+1) \sisos^2
%		 \nonumber \\ &&
		 ~+~ (2S+1)(2T-1) \rhoisov^{~2}
		 + (2S-1)(2T-1)\sisov^{~2}
	\big] \nonumber\\
	&& +~ \big[ \ctau_0 + (4S-3) \csT_0 + (4T-3) \ctau_1
		+ (4S-3)(4T-3)\csT_1 \big] \nonumber \\ &&
		~\times~ \frac{1}{16}
	\big[ (2S+1)(2T+1)\rhoisos\tauisos + (2S-1)(2T+1)\sisos\cdot\Tisos
%	\nonumber \\ &&
		~+~ (2S+1)(2T-1)\rhoisov\tauisov
		+ (2S-1)(2T-1)\sisov\cdot\Tisov
	\big]. \nonumber \\
\end{wideeqnarray}

%%%%%%%%%%%%%%%%%%%%%%%%%%%%%%%%%%%%%%%%%%%%%%%%%%%%%%%%%%%%%%%%%%%%%%%%%%%%%%%

\section{Particle-hole residual interaction}

%%%%%%%%%%%%%%%%%%%%%%%%%%%%%%%%%%%%%%%%%%%%%%%%%%%%%%%%%%%%%%%%%%%%%%%%%%%%%%%

With the definition of densities given in Eqs.~(\ref{eq:densdefbegin})-
(\ref{eq:densdefend}) the particle-hole residual interaction is obtained
through
\begin{eqnarray}
&&\langle \xvec'_1 \sigma'_1 q'_1,  \xvec'_2 \sigma'_2 q'_2 \vert
V_\ph \vert \xvec_1 \sigma_1 q_1,  \xvec_2 \sigma_2 q_2 \rangle \nonumber \\
	&& ~=~ \frac{\delta^2 \CE}%
	{\delta \rho(\xvec_1 \sigma_1 q_1, \xvec'_1 \sigma'_1 q'_1)~
	 \delta \rho(\xvec_2 \sigma_2 q_2, \xvec'_2 \sigma'_2 q'_2)},
\end{eqnarray}
which, for the central, spin-scalar, isoscalar part of the functional,
\emph{i.~e.~}
\begin{eqnarray}
\CH^\isv &=& \crho_0(\rhoisos)~\rhoisos^2 + \cdrho_0~\rhoisos \Delta~\rhoisos
    + \ctau_0~(\rhoisos~\tauisos - \jisos^2), \nb \\
\end{eqnarray}
(the generalization to the full central part, omitted here for the sake of
brevity, being straightforward), reads

\begin{eqnarray}
V_\ph &=& 2 \crhoz_0 + \sum_{i=1}^2 \crhoi_0
		\left(\frac{i}{3}+2\right)\left(\frac{i}{3}+1\right) \rhoit
       \nonumber \\
  &&+ \ctau_0 \Big( \gradvec'_1\cdot\gradvec_1 + \gradvec'_2\cdot\gradvec_2
  \nonumber \\
	&&~+ \frac{1}{2}
		(\gradvec'_1 - \gradvec_1)\cdot(\gradvec'_2 - \gradvec_2) \Big)
	\nonumber \\
	&&+ \cdrho_0
	  \Big( (\gradvec'^2_1 + 2\gradvec'_1\cdot\gradvec_1 + \gradvec_1^2)
	\nonumber \\
     &&~+ (\gradvec'^2_2 + 2\gradvec'_2\cdot\gradvec_2 +\gradvec_2^2) \Big).
\end{eqnarray}
When computing the momentum-space matrix element, \citeeqdot{vphkspace}, one
makes the substitutions $\gradvec_1 = i (\qvec_1 + \qvec)$,
$\gradvec_2 = i \qvec_2$, $\gradvec'_1 = -i \qvec_1$ and
$\gradvec'_2 = -i (\qvec_2 + \qvec)$ (with $\hbar=1$),
which yields expressions (\ref{eq:wfuncsdefbegin})-(\ref{eq:wfuncsdefend}).

%%%%%%%%%%%%%%%%%%%%%%%%%%%%%%%%%%%%%%%%%%%%%%%%%%%%%%%%%%%%%%%%%%%%%%%%%%%%%%%

\bibliography{meff}

\newpage

%%%%%%%%%%%%%%%%%%%%%%%%%%%%%%%%%%%%%%%%%%%%%%%%%%%%%%%%%%%%%%%%%%%%%%%%%%%%%%%

\end{document}